\documentclass[12pt,preprint]{aastex}
\usepackage{graphicx}
\usepackage{emulateapj5}
\usepackage{apjfonts}
\usepackage{epsfig}
\usepackage{natbib}
\usepackage{amsmath}
\usepackage[usenames]{color}
\usepackage{CJK}
\def\gtrsim{\mathrel{\hbox{\rlap{\hbox{\lower4pt\hbox{$\sim$}}}\hbox{\raise2pt\hbox{$>$}}}}}

\newcommand{\kms}{km~s\ensuremath{^{-1}}}

\newcommand{\lamout}{\ensuremath{\lambda_{\rm out}}}
\newcommand{\lamre}{\ensuremath{\lambda_{\rm R_e}}}

\newcommand{\msun}{\ensuremath{M_{\odot}}}

\newcommand{\sers}{S{\'e}rsic}

\def\lax{{$\mathrel{\hbox{\rlap{\hbox{\lower4pt\hbox{$\sim$}}}\hbox{$<$}}}$}}
\def\gax{{$\mathrel{\hbox{\rlap{\hbox{\lower4pt\hbox{$\sim$}}}\hbox{$>$}}}$}}

\begin{document}

\title{SDSS-IV MaNGA: Uncovering the Angular Momentum Content of Central and Satellite Early-type Galaxies}

\author{J. E. Greene\altaffilmark{1}, A. Leauthaud\altaffilmark{2}, E. Emsellem\altaffilmark{3}, J. Ge\altaffilmark{4}, A. Arag\'on-Salamanca\altaffilmark{5}, J. Greco\altaffilmark{1}, Y.-T. Lin\altaffilmark{6}, S. Mao\altaffilmark{4,7,8}, K. Masters\altaffilmark{9}, M. Merrifield\altaffilmark{5}, S. More\altaffilmark{10}, N. Okabe\altaffilmark{11,12,13}, D.~P.~Schneider\altaffilmark{14,15}, D. Thomas\altaffilmark{9}, D.~A.~Wake\altaffilmark{16,17}, K. Pan\altaffilmark{18}, D. Bizyaev\altaffilmark{18,19}, D. Oravetz\altaffilmark{18},
A. Simmons\altaffilmark{18}, R. Yan\altaffilmark{19}, F. van den Bosch\altaffilmark{20}}

\altaffiltext{1}{Department of Astrophysics, Princeton University, Princeton, NJ 08540, USA}
\altaffiltext{2}{Department of Astronomy and Astrophysics, University of California, Santa Cruz, 1156 High Street, Santa Cruz, CA 95064, USA}
\altaffiltext{3}{European Southern Observatory, Karl-Schwarzschild-Str. 2, D-85741 Garching, Germany}
\altaffiltext{4}{National Astronomical Observatories, Chinese Academy of Sciences, 20A Datun Road, Chaoyang District, Beijing 100012, China}
\altaffiltext{5}{School of Physics and Astronomy, The University of Nottingham, University Park, Nottingham, NG7 2RD, UK}
\altaffiltext{6}{Institute of Astronomy and Astrophysics, Academia Sinica, Taipei 10617, Taiwan}
\altaffiltext{7}{Physics Department and Tsinghua Centre for Astrophysics, Tsinghua University, Beijing 100084, China}
\altaffiltext{8}{Jodrell Bank Centre for Astrophysics, School of Physics and Astronomy, The University of Manchester, Oxford Road, Manchester M13 9PL, UK}
\altaffiltext{9}{Institute of Cosmology and Gravitation, University of Portsmouth, Dennis Sciama Building, Burnaby Road, Portsmouth PO1 3FX, UK; South East Physics Network}
\altaffiltext{10}{Kavli Institute for the Physics and Mathematics of the Universe (WPI), Tokyo Institutes for Advanced Study, The University of Tokyo, 5-1-5 Kashiwanoha, Kashiwa-shi, Chiba, 277-8583, Japan}
\altaffiltext{11}{Department of Physical Science, Hiroshima University, 1-3-1 Kagamiyama, Higashi-Hiroshima, Hiroshima 739-8526, Japan}
\altaffiltext{12}{Hiroshima Astrophysical Science Center, Hiroshima University, Higashi-Hiroshima, Kagamiyama 1-3-1, 739-8526, Japan}
\altaffiltext{13}{Core Research for Energetic Universe, Hiroshima University, 1-3-1, Kagamiyama, Higashi-Hiroshima, Hiroshima 739-8526, Japan}
\altaffiltext{14}{Department of Astronomy and Astrophysics, The Pennsylvania State University,University Park, PA 16802}
\altaffiltext{15}{Institute for Gravitation and the Cosmos, The Pennsylvania State University, University Park, PA 16802}
\altaffiltext{16}{School of Physical Sciences, The Open University, Milton Keynes,  MK7 6AA, UK}
\altaffiltext{17}{Department of Physics, University of North Carolina, Asheville, NC 28804, USA}
\altaffiltext{18}{Apache Point Observatory and New Mexico State
University, P.O. Box 59, Sunspot, NM, 88349-0059, USA}
\altaffiltext{19}{Sternberg Astronomical Institute, Moscow State
University, Moscow}
\altaffiltext{19}{Department of Physics and Astronomy, University of Kentucky, 505 Rose Street, Lexington, KY 40506-0057, USA}
\altaffiltext{20}{Department of Astronomy, Yale University, PO. Box 208101, 
  New Haven, CT 06520-8101}

\maketitle

\begin{abstract}
    We study 379 central and 159 satellite early-type galaxies with two-dimensional kinematics from the integral-field survey Mapping Nearby Galaxies at APO (MaNGA) to determine how their angular momentum content depends on stellar and halo mass. Using the \citet{yangetal2007} group catalog, we identify central and satellite galaxies in groups with halo masses in the range $10^{12.5}~h^{-1}~M_{\odot}< M_{200b} < 10^{15}~h^{-1}~M_{\odot}$. As in previous work, we see a sharp dependence on stellar mass, in the sense that $\sim 70\%$ of galaxies with stellar mass $M_* > 10^{11}~h^{-2}~M_{\odot}$ tend to have very little rotation, while nearly all galaxies at lower mass show some net rotation.  The $\sim 30\%$ of high-mass galaxies that have significant rotation do not stand out in other galaxy properties except for a higher incidence of ionized gas emission.  Our data are consistent with recent simulation results suggesting that major merging and gas accretion have more impact on the rotational support of lower-mass galaxies. When carefully matching the stellar mass distributions, we find no residual differences in angular momentum content between satellite and central galaxies at the 20\% level. Similarly, at fixed mass, galaxies have consistent rotation properties across a wide range of halo mass. However, we find that errors in classification of centrals and satellites with group finders systematically lowers differences between satellite and central galaxies at a level that is comparable to current measurement uncertainties. To improve constraints, the impact of group finding methods will have to be forward modeled via mock catalogs.
\end{abstract}

\section{Introduction}

Integral field spectroscopic (IFS) surveys have provided an exciting new window into the formation histories of massive galaxies \citep[e.g.,][]{cappellari2016}. We have long understood that rotational support $V/\sigma$ (with $V$ the maximum radial velocity, and $\sigma$ the stellar velocity dispersion) is a function of internal galaxy properties, with lower-mass ellipticals and S0s (early-type galaxies) having more rotation than the most massive elliptical galaxies \citep[e.g.,][]{binney1978,daviesetal1983,franxillingworth1990,benderetal1989,kormendybender1996,kormendybender2009}. IFS studies have definitively shown that $\sim L^*$ early-type galaxies typically have some net rotation \citep{emsellemetal2007}, while the most massive elliptical galaxies tend to have little to no net rotation \citep[e.g.,][]{emsellemetal2011,houghtonetal2013,scottetal2014,raskuttietal2014,vandesandeetal2017,vealeetal2017a,vealeetal2017b,olivaetal2017,broughetal2017}. 

In the era of IFS surveys, the classic ratio of $V/\sigma$ that quantifies the level of rotation in galaxies has been replaced by a light-weighted two-dimensional analog, $\lambda$. We adopt the definition of $\lambda_R$ from \citet[][]{emsellemetal2007}, inspired by the two-dimensional modeling of $V / \sigma$ from first principles as described in \citet{binney2005}, with $V$ and $\sigma$ the locally measured values, $R$ the projected galactocentric distance, and brackets representing flux weighting: 

\begin{equation}
\lambda_R = \langle R | V | \rangle/\langle R \sqrt{V^2 + \sigma^2} \rangle. 
\end{equation}

Note that $\lambda_R$ is a cumulative measurement out to radius $R$, which has been shown to be a robust proxy for the spin parameter \citep[see][]{emsellemetal2007,jesseitetal2009}. Early surveys did not cover sufficient volume to study the importance of extrinsic factors (such as local galaxy density) on the galaxy angular momentum content. The ATLAS$^{\rm 3D}$ survey \citep{cappellarietal2011},  volume-limited to 42 Mpc, includes few slowly rotating galaxies and contains only one dense environment (the Virgo Cluster). In principle, there are many reasons to believe that the location of a galaxy within its halo may impact its angular momentum. For instance, if a series of minor mergers can lead to loss of net angular momentum \citep[e.g.,][]{naabetal2014} then central galaxies, that likely experience enhanced minor merging due to their location at the center of the potential, may be expected to preferentially lack rotation. The halo mass could also matter, since galaxies in more massive halos may have experienced more merging, even at fixed $M_*$. 

Ongoing large-scale IFS surveys, including CALIFA \citep{sanchezetal2012}, SAMI \citep{croometal2012,bryantetal2015}, MASSIVE \citep{maetal2014}, and MaNGA \citep[][]{bundyetal2015} provide the larger volumes that are needed to simultaneously control for galaxy properties, such as stellar mass, along with large-scale environment. They also provide a larger baseline in halo mass, providing the leverage to examine trends in stellar mass $M_*$ and halo mass $M_{200b}$ simultaneously, as well as separate central and satellite galaxies. Thanks to the Sloan Digital Sky Survey \citep[SDSS;][]{yorketal2000}, we now have the sample size to investigate all of these important parameters in setting the angular momentum content of galaxies. In this paper, we will use spatially resolved spectroscopy from the MaNGA survey \citep[part of SDSS IV;][]{blantonetal2017} to address the relationship between internal galaxy properties, large-scale environment, and angular momentum, $\lambda_R$. We address the possible correlation between $\lambda_R$ and local overdensity in a companion paper.

This paper proceeds as follows. In \S \ref{sec:data}, we discuss the MaNGA survey and our galaxy sample. We present the kinematic measurements in \S \ref{sec:analysis}, and examine the results in \S \ref{sec:results}. We conclude in \S \ref{sec:summary}. For consistency with our group catalog from \citet{yangetal2007} (\S 2), we assume a flat $\Lambda$CDM cosmology with $\Omega_{\rm m}=0.238$, $\Omega_\Lambda=0.762$, H$_0=$100~h$^{-1}$~km~s$^{-1}$~Mpc$^{-1}$. Halo masses are defined as $M_{200b}\equiv M(<r_{200b})=200\bar{\rho} \frac{4}{3}\pi R_{200b}^3$ where $R_{200b}$ is the radius at which the mean interior density is equal to 200 times the mean matter density ($\bar{\rho}$) and the `b' indicates background (rather than critical) density. Stellar mass is denoted $M_{*}$ and has been derived using a \citet{chabrier2003} Initial Mass Function (IMF). For consistency with our adopted stellar mass measurements, we assume $h=1$.

\section{Data and Sample}
\label{sec:data}

\begin{figure*}
\vbox{ 
\vskip 0mm
\hskip +20mm
\includegraphics[width=0.65\textwidth]{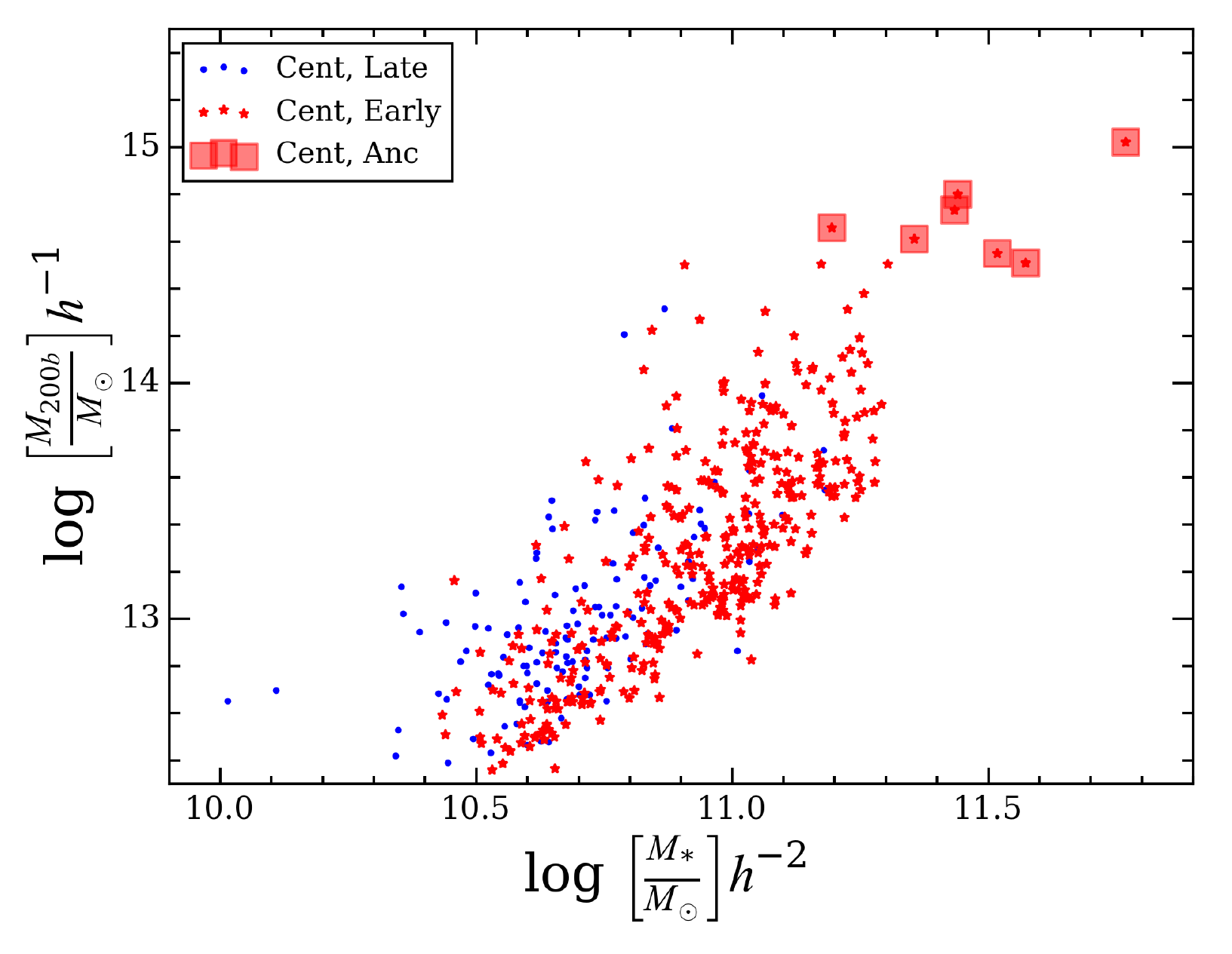}
}
\vskip -0mm
\figcaption[]{
Distribution of $M_{200b}$ and $M_*$ for the central galaxies in our sample. For context, we show early-type central galaxies with red stars and late-type centrals with blue circles. Late-type galaxies are identified as those with spiral structure in the SDSS images (see \S 2.3.1). However, throughout the manuscript we focus on early-type centrals. Our seven Ancillary program targets (red squares) clearly fill in the highest stellar and halo masses in our sample.
\label{fig:bcgsample}}
\end{figure*}

\subsection{The MaNGA Survey}
 
The aim of MaNGA, one of three core SDSS IV projects, is to obtain integral-field spectroscopy of 10,000 nearby galaxies. Like previous SDSS surveys, MaNGA utilizes the 2.5m Sloan Foundation Telescope \citep{gunnetal2006} and the BOSS spectrographs \citep{smeeetal2013}. Unlike previous SDSS surveys, the MaNGA survey groups individual fibers into 17 hexagonal fiber bundles to perform a multi-object IFS survey \citep{droryetal2015}. Each fiber has a diameter of 2\arcsec, while the bundles range in diameter from 12 to 32\arcsec\ with a 56\% filling factor. The two dual BOSS spectrographs cover the wide wavelength range of 3600-10,300\AA\ while maintaining a spectral resolution of $\sigma_r \approx 70$~\kms, appropriate for galaxy studies. Careful spectrophotometry yields a relative calibration accurate to a few percent \citep{yanetal2016a}. The survey design is described in \citet{yanetal2016b}, the observing strategy in \citet{lawetal2015}, and the data reduction pipeline in \citet{lawetal2016}. The MaNGA team has also developed useful tools for data visualization and vetting \citep{cherinkaetal2017}.

The MaNGA sample is selected from an enhanced version of the NASA-Sloan Atlas (NSA; Blanton M.; http://www.nsatlas.org) with redshifts primarily taken from the SDSS DR7 MAIN galaxy sample \citep{abazajianetal2009} and photometry from reprocessed SDSS imaging. The sample is built in $i$-band absolute magnitude ($M_i$)-complete shells, with more luminous galaxies observed in more distant $M_i$ shells such that the spatial resolution (in terms of $R_e$) is roughly constant across the sample \citep[][]{yanetal2016b,wakeetal2017}. The MaNGA {\it Primary} sample, which forms $\sim 50\%$ of the total sample, is selected such that $80\%$ of the galaxies in each $M_i$ shell can be covered to $1.5 R_e$ by the largest MaNGA IFU. There is also a {\it Secondary} sample, accounting for $\sim 40\%$ of the targets, selected such that $80\%$ of these galaxies are covered to $2.5 R_e$. The remainder of the MaNGA main sample is the {\it Color-enhanced} supplement, which fills in poorly-covered regions of the color-magnitude diagram (e.g., faint red galaxies or the most luminous blue galaxies) and, like the Primary sample, is covered to $1.5 R_e$. 

Given this sample construction, one must re-weight the galaxy distribution according to the volume in each shell to construct volume-limited samples. Whenever possible, we apply these weights before drawing conclusions. In this work we take a conservative approach and consider the Primary and Secondary samples only, excluding the Color-enhanced supplement. We make this choice since the addition of the color term in the Color-enhanced selection potentially increases the uncertainty in volume weights for these galaxies. 

\subsection{Ancillary Program}
 
The MaNGA survey awarded a small fraction of fiber-bundles to ``Ancillary'' Programs that can help to boost survey efficiency by serving as filler targets and also enhance the science output of MaNGA with a small time investment. We were awarded an Ancillary program to augment the number of central galaxies in the most massive halos available within the local volume targeted by MaNGA. Here we describe our target selection, along with a brief description of the group catalog that our selection is based upon.

While there are several samples of central galaxies in the literature, building an uncontaminated sample of central galaxies spanning a range of halo masses is a non-trivial exercise. Different cluster finders and central galaxy selections, automated and visual, often disagree. We base our target selection on the \citet[][Y07]{yangetal2007} cluster catalog updated to DR7, created from the SDSS DR7 New York University Value Added Catalog \citep[VAGC;][]{blantonetal2005}, a spectroscopic galaxy catalog.

\begin{figure*}
\hskip 30mm
\vbox{
\vbox{ 
\vskip -1mm
\includegraphics[width=0.6\textwidth]{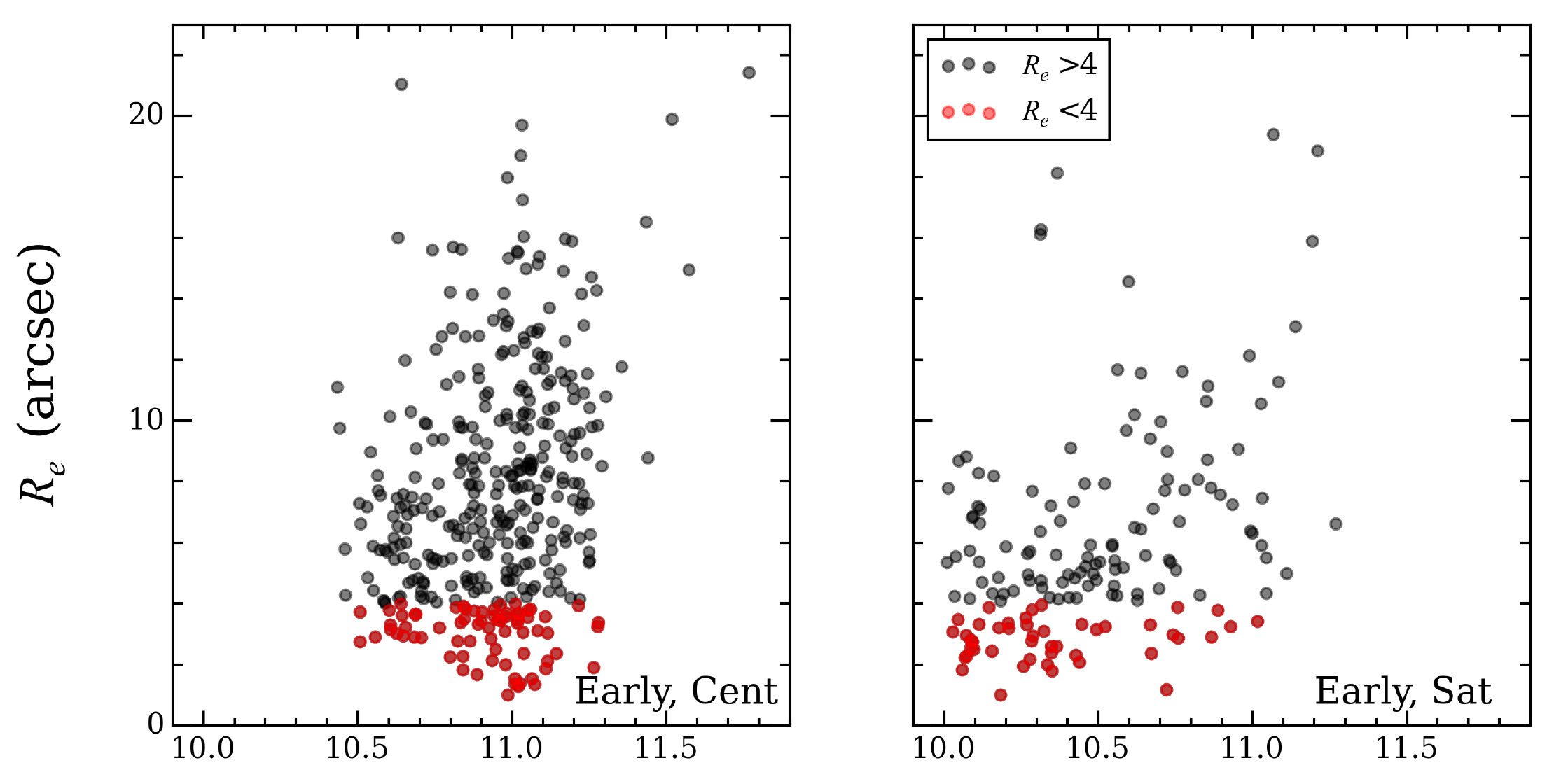}
}
\vbox{
\includegraphics[width=0.6\textwidth]{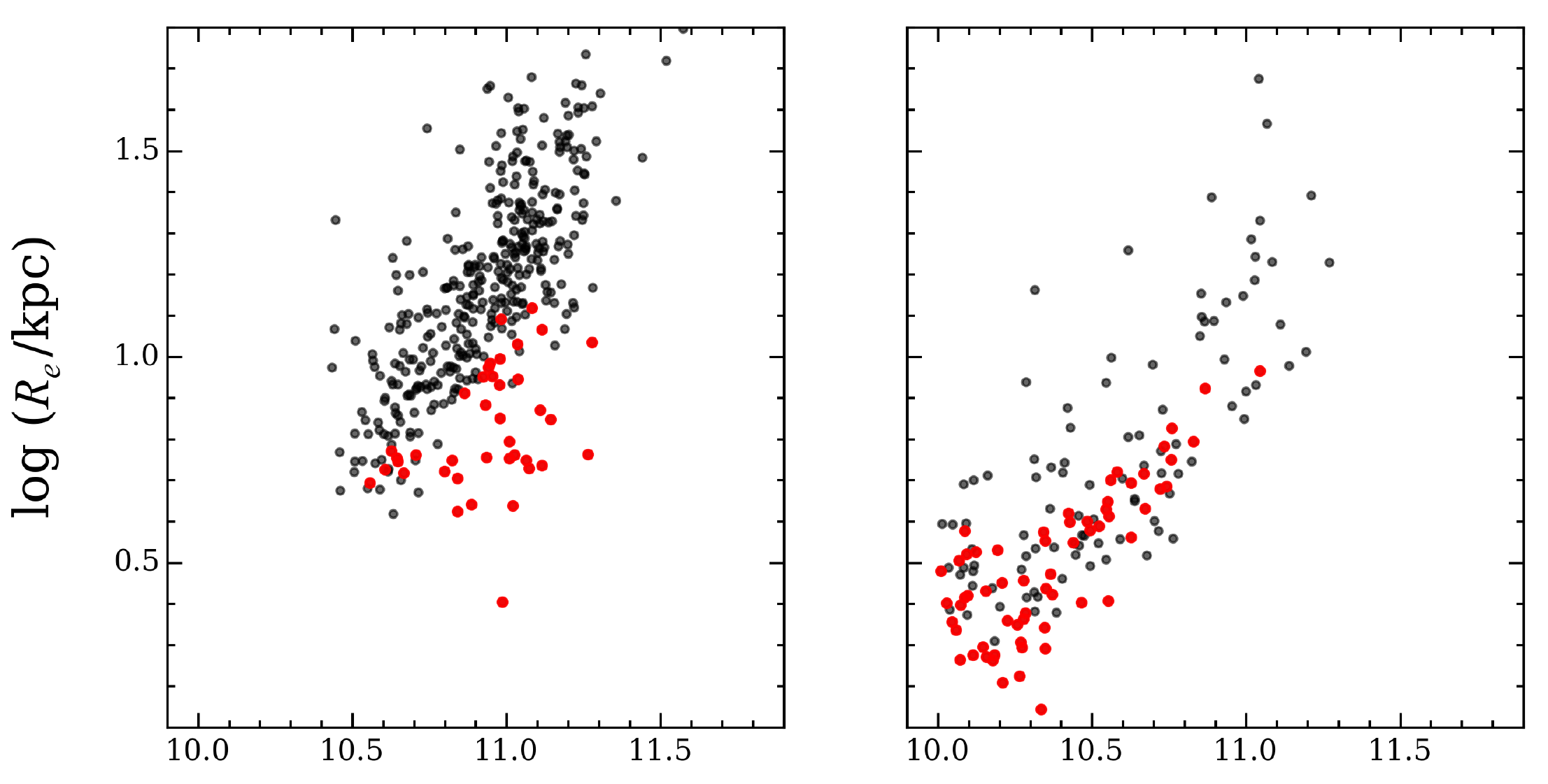}
}
\vbox{
\includegraphics[width=0.6\textwidth]{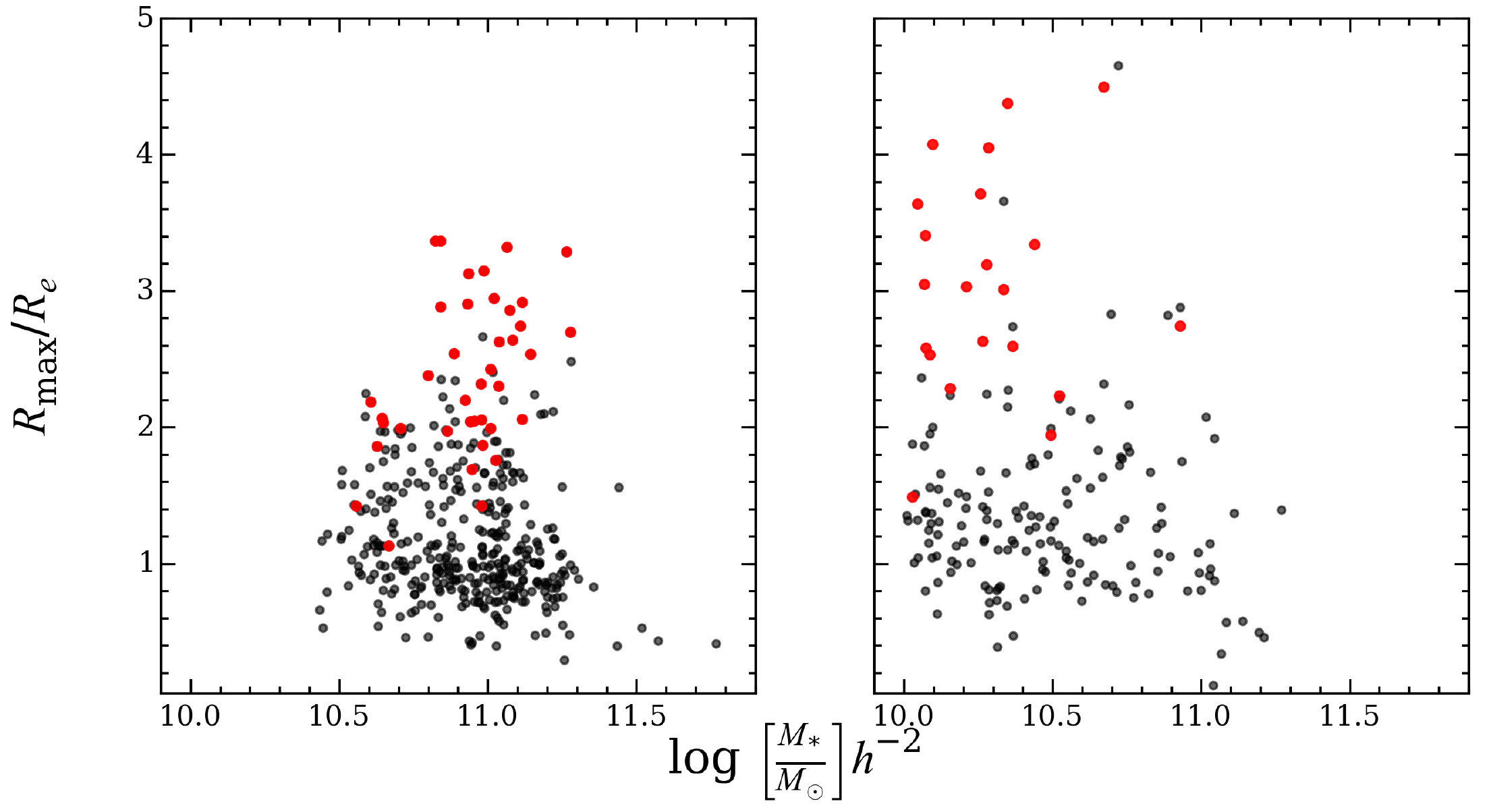}
}}
\vskip -0mm
\figcaption[]{
{\it Top Row}: The relationship between stellar mass and angular size for the early-type sample only. 
While a large fraction of the galaxies are well-resolved (black), there is a tail of smaller galaxies (red) that are not. We highlight galaxies with $R_e <4$\arcsec, which is 2.5 times the typical PSF radius of 1.5\arcsec \citep{yanetal2016b}. The limited spatial resolution directly impacts our ability to measure $\lambda_R$ (\S 3.2). When considering primary and secondary samples separately, there is not a mass dependence in resolved fraction over this stellar mass range due to the construction of the MaNGA sample (\S 3.1.1).
{\it Middle Row}: The relationship between stellar mass and galaxy size ($R_e$ in kpc). Red and black symbols as above. Only for the most compact galaxies in the lowest stellar mass bin do we have a large fraction of unresolved galaxies.
{\it Bottom Row}: Same as the middle row, but in $R_\mathrm{max}/R_e$ units, where $R_\mathrm{max}$ is the largest $R_\mathrm{Area}$ where we have a $\lambda_R$ measurement. Red and black symbols as above. These panels summarize our radial coverage.
\label{fig:mass_size}}
\end{figure*}

Y07 use an iterative, adaptive group finder to assign galaxies to halos. In short, they first use a friends-of-friends algorithm \citep[e.g.,][]{davisetal1985} to identify potential groups. Each group is assigned a characteristic luminosity, defined as the combined $r$-band luminosity of all group members with $M_r-5 \log(h) \leq-19.5$~mag where $M_r$ is the absolute Galactic-extinction corrected $r$-band luminosity, $K$-corrected to $z = 0.1$. Roughly speaking, the stellar mass comprises 1\% of the total mass. This characteristic luminosity is used to assign halo masses to groups and to refine the group identification using an iterative method. Using a mock catalog, Y07 estimate the scatter in the assigned halo masses to be of order $\sigma_{Q}\sim0.35$ dex for groups with 10$^{13}h^{-1}M_{\odot}<M_{200b}<10^{14.5}h^{-1} \, M_{\odot}$, where $\sigma_{Q}$ is the standard deviation of $Q=[\log(M_L)-\log(M_{200b})]/\sqrt{2}$, with $M_L$ the halo mass inferred from the group luminosity. They report that the overall scatter is dominated by intrinsic scatter between halo mass and $M_L$, while the details of the group finder (interlopers, incompleteness effects), are a relatively small component. \citet{campbelletal2015} demonstrate that in general it is possible to extract meaningful physical correlations from Y07 as a function of color, stellar, and halo mass despite misidentifications and errors in halo mass \citep[although see also][]{tekluetal2017}.

\begin{figure*}
\hskip 25mm
\vbox{ 
\includegraphics[width=0.36\textwidth]{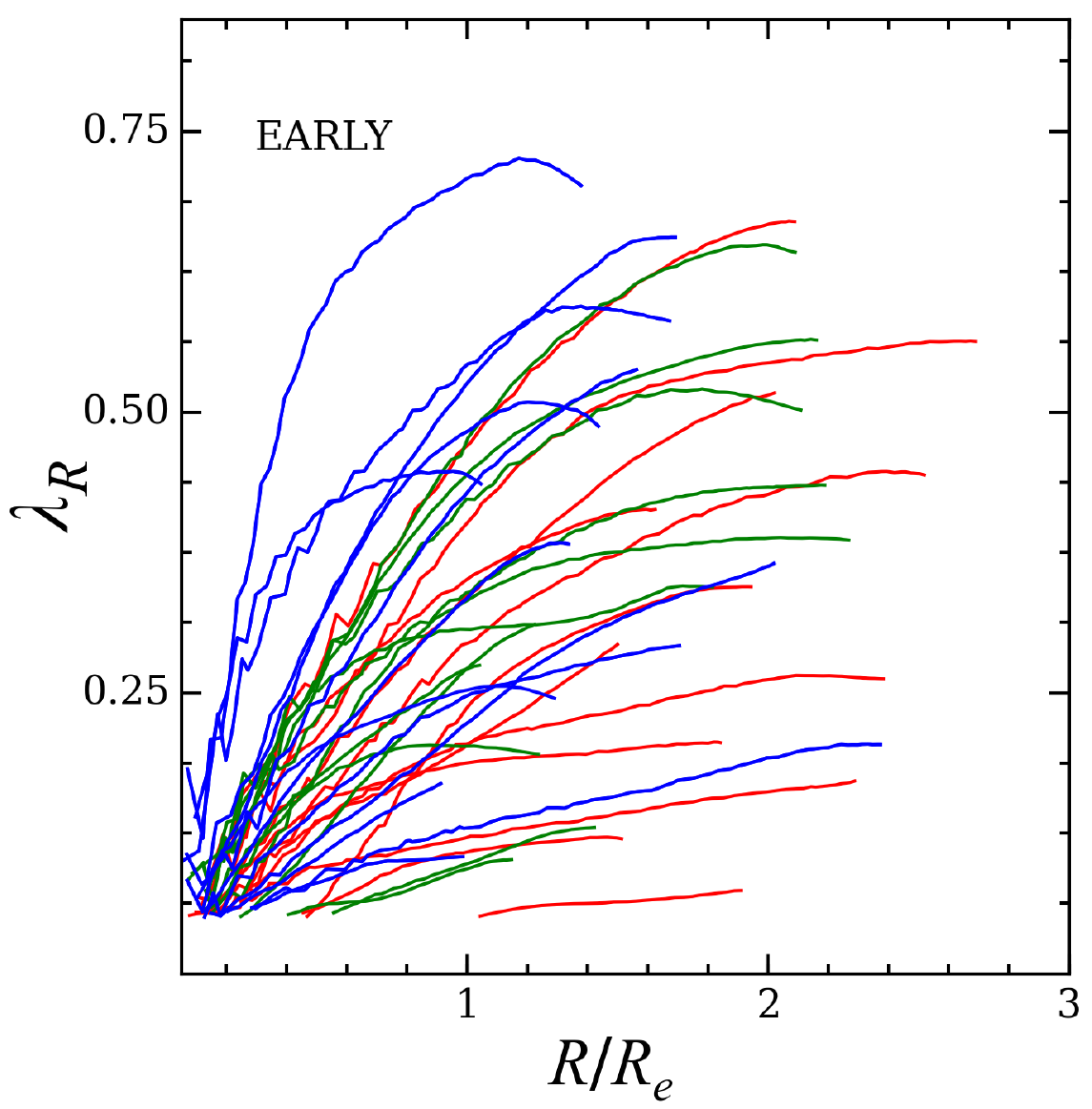}
\includegraphics[width=0.35\textwidth]{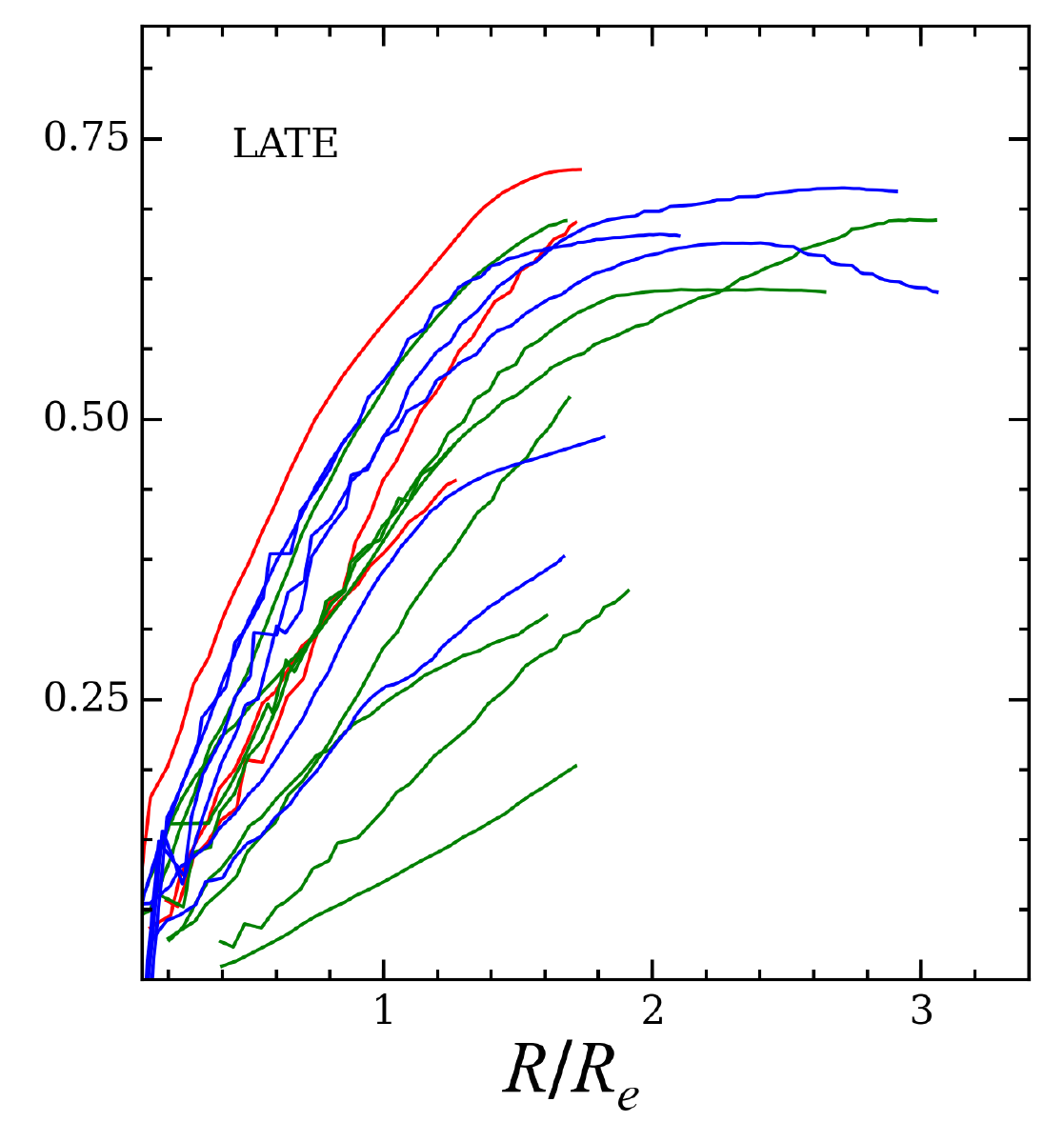}
}
\vskip -0mm
\figcaption[]{
Measurements of $\lambda_R$ as a function of radius (measured as $R_\mathrm{Area}$ defined in \S 3.2) for a set of representative galaxies in three stellar mass bins centered at $10^{10.5}, 10^{10.75}, \mathrm{and} 10^{11}~h^{-2}~M_{\odot}$ (blue, green, red lines), respectively. Note that $\lambda_R$ is cumulative to radius $R$, as is customary in the literature \citep[although see][]{raskuttietal2014}. We separate the systems into early-type (left), and late-type (right). While we show the late-type galaxies here for comparison, throughout the manuscript we focus on early-type centrals. Any changes in $\lambda_R$ at larger radii do not change the slow or fast rotator designation.
\label{fig:lamradius}}
\end{figure*}

We adopt the Y07 modelC catalog, which uses the SDSS model magnitudes and includes redshifts from SDSS and the 2dF Galaxy Redshift Survey \citep[][]{collessetal2001} and nearest neighbors from the VAGC. We identify central galaxies as the most luminous (in $r$-band) galaxy, as noted in the Y07 imodelC\_1 catalog. We adopt the Y07 group halo mass based on the total luminosity ranking of the groups, which is in the modelC\_group file in the Y07 catalog. To build our sample, we avoid edges of the catalog by applying $grp\_f\_edge>0.6$ as recommended by Y07, and $0.02<z_{\rm group}<0.15$, with the lower limit reflecting the Y07 limits and the upper limit set to match the MaNGA sample. 
 
Because massive halos are rare, the default Primary and Secondary MaNGA catalogs do not contain many central galaxies in high-mass halos (Figure \ref{fig:bcgsample}). To address this lack, in our ancillary program we select central galaxies in halos more massive than $M_{200b}=10^{13.75}~h^{-1}~M_{\odot}$, dividing our sample into four halo mass bins. We construct the ancillary sample such that in each halo bin, we add sufficient galaxies to enable stacked stellar population gradient measurements in each halo bin for two bins in stellar velocity dispersion. In practice, we insist on a stacked S/N of 50 at a radius of $1.25-1.75 R_e$. In total, the ancillary program aims to observe $\sim 50$ additional central galaxies to add to the Primary and Secondary MaNGA observations, of which seven are included in this paper.

\hskip +0mm
\vbox{
\vbox{ 
\vskip -1mm
\includegraphics[width=0.45\textwidth]{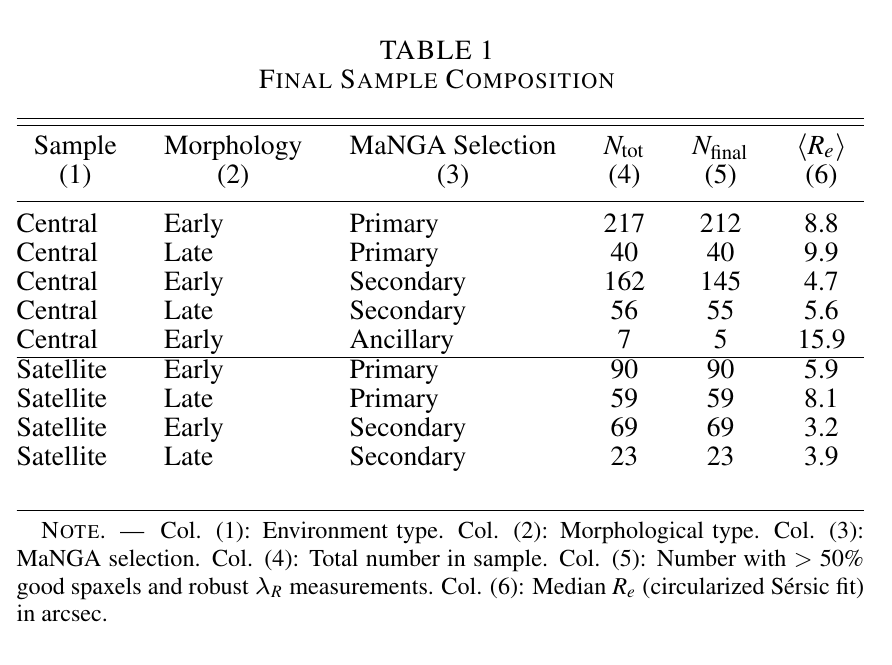}
}}
\vskip 5mm

\subsection{Galaxy Sample}

We now turn to the properties of the entire sample of galaxies considered here. We are working with the MaNGA Data Reduction Pipeline version 2.0.1 sample (MaNGA Product Launch 5; MPL5; K.~Westfall et al.\ in preparation). An initial set of central galaxies are selected from the Y07 catalog and are defined as the most luminous galaxy among the group members. Two coauthors performed visual inspection in the $r-$band of all of the Y07 groups with $M_{200b}>10^{14}~h^{-1}~M_{\sun}$, the halo-mass range that we focus on for the ancillary sample. In doing these visual checks, we both ensure that the chosen overdensity exists and check the validity of the choice of central galaxy. Based on the visual inspection, we judge that the Y07 algorithm overall selected visually reasonable clusters and central galaxy candidates. We apply a stellar mass cut $M_* > 10^{10}~h^{-2}~M_{\odot}$ (in practice, 95\% of central galaxies have stellar masses $M_*> 10^{10.5}~h^{-2}~M_{\odot}$) and a halo mass cut $M_{200b}> 10^{12.5}~h^{-1}~M_{\odot}$. As shown in \citet{yangetal2009}, the groups are quite incomplete below this halo mass in the redshift range of interest. As a result, the majority of our central galaxies have stellar masses $M_*>10^{10.5}~h^{-2}~M_{\odot}$. Satellite galaxies are those in the Y07 catalog that are not central galaxies. Of course, the satellite galaxies extend to much lower stellar masses. However, our primary goal here is to compare the satellite and central galaxy populations. Furthermore, at low stellar mass, the early-type galaxies in MaNGA are overwhelmingly unresolved (see also Appendix B). Therefore, we apply a stellar mass cut to the satellite galaxies of $M_*> 10^{10}~h^{-2}~M_{\odot}$. 

\begin{figure*}
\vbox{ 
\vskip -2mm
\hskip +10mm
\includegraphics[width=0.85\textwidth]{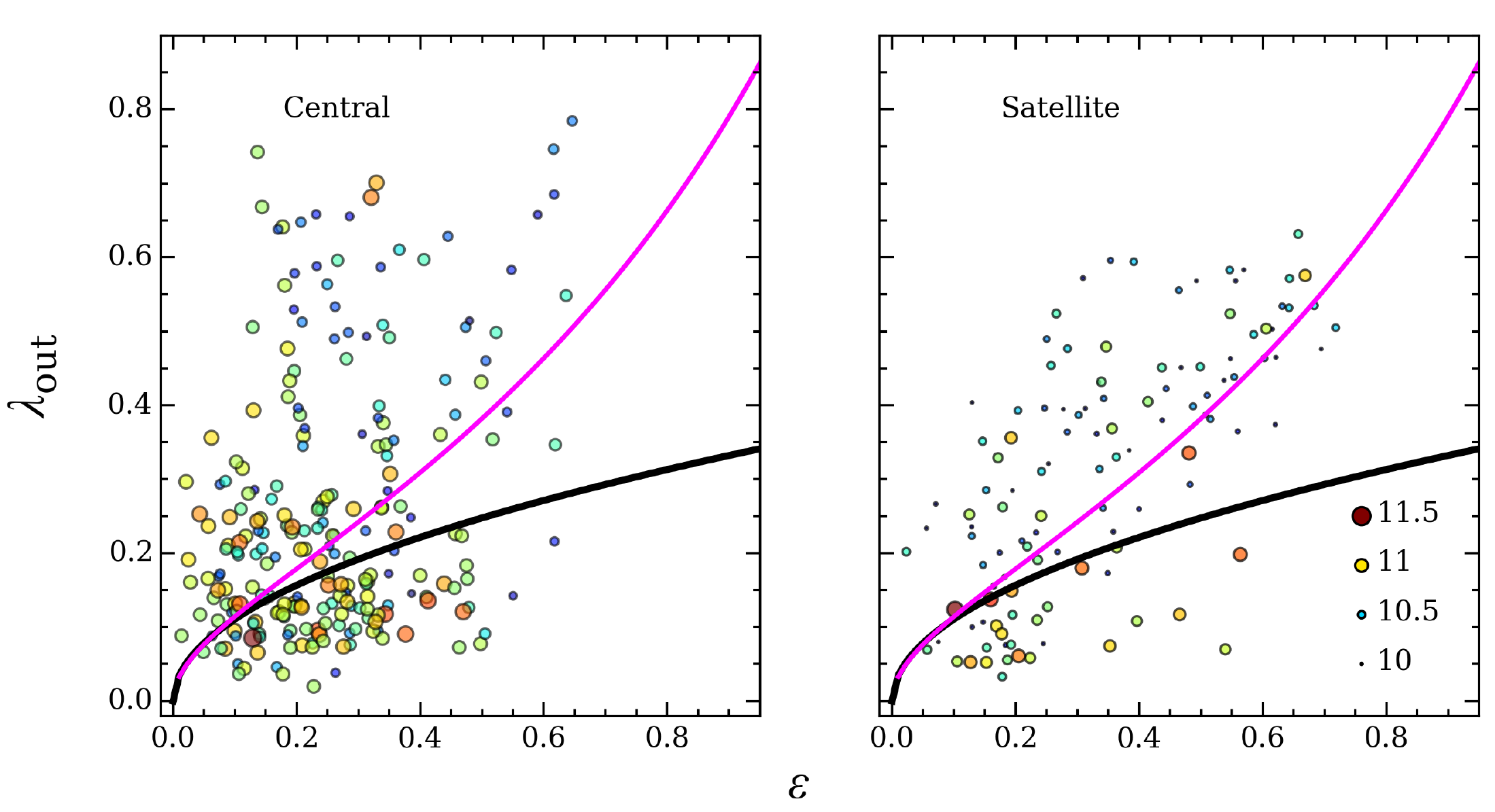}
}
\vskip -0mm
\figcaption[]{
Angular momentum content of the early-type central (left) and satellite (right) galaxies as traced by \lamout, plotted as a function of the isophote flattening~$\epsilon$. There is a bias against $\lambda = 0$ due to errors present in the $V$ measurements. Stellar mass is indicated by size and color of the symbol. The thick black line indicates the empirical division between fast (above the line) and slow (below the line) rotators. Here we use an empirical division of $0.35 \sqrt{\epsilon}$ adapted from \citet[][]{emsellemetal2007}, but scaled from $R_e$ to $1.5 R_e$ to match the typical radial extent of our galaxies. Depending on the anisotropy and inclination angle, galaxies will have a different relationship between $\epsilon$, the observed value of $\lambda_R$, and the edge-on value. The magenta line represents an analytic model edge-on galaxy in which anisotropy is proportional to ellipticity \citep[with $\beta=0.7$ as in ][]{cappellarietal2007}. Note that low-$\lambda$ galaxies with $\epsilon>0.4$ are quite rare, and may all be explained by more exotic kinematics \citep[\S 3.2.2][]{cappellarietal2012}. 
\label{fig:lam_eps}}
\end{figure*}

We adopt measured properties (e.g., stellar mass, galaxy radius, redshift) from the MaNGA source catalog, which in turn is based on version v1\_0\_1 of the NSA. The galaxy magnitudes are based on elliptical Petrosian apertures, measured as Petrosian magnitudes \citep{petrosian1976,blantonetal2001} but using elliptical apertures\footnote{http://www.sdss.org/dr13/manga/manga-target-selection/nsa/}. The stellar masses are derived using the k-correct code \citep{blantonetal2003}, which fits spectral energy distributions to the elliptical Petrosian magnitudes to derive the mass-to-light ratio. A \citet{chabrier2003} Initial Mass Function is assumed.

Typically $\lambda_R$ is measured in a fixed aperture (often $R_e$), so that all galaxies can roughly be on the same footing. There are two complications to this approach for our sample. One is that the galaxies are typically poorly resolved spatially at $R_e$, which compromises the $\lambda_R$ measurements (see Appendix B). The other issue is that it is notoriously difficult to measure a uniform effective radius \citep[e.g.,][]{kormendyetal2009}. In the end, we therefore adopt a measurement of $\lambda_R$ that is not directly tied to $R_e$. Nevertheless, we must adopt some measure of size.

There are two sizes tabulated by the NSA that we consider here. The one used by the MaNGA team to define the sample is the elliptical Petrosian radius. The other possibility within the NSA is the $R_e$ provided by a single-\sers\ fit. We adopt the aspect ratio (B/A) and position angle (PA) derived from the parametric \sers\ fit because they are PSF-corrected. Indeed, when we compare B/A derived from the Petrosian and \sers\ fits we find clear evidence that the PSF-correction makes a difference, since the \sers\ B/A is typically 10\% smaller than the Petrosian value. To be consistent between the size measurement $R_e$ and the ellipticity and PA measurements, we adopt the circularized $R_e$ from the \sers\ fit. All $R_e$ measurements use the \sers\ fits. This measurement is roughly 30\% larger than the elliptical Petrosian measurements used to define the MaNGA targets. Again, we emphasize that this decision does not impact the final $\lambda_R$ measurements, but the circularized \sers-derived $R_e$ is used as a benchmark throughout the paper.

\subsubsection{Galaxy Morphologies}

Many of the central galaxies are late-type (spiral) galaxies. Late-type galaxies tend to have high $\lambda_R$ values, and our main goal is to investigate the distribution in $\lambda_R$ for the early-type galaxies. We have visually classified galaxies into those with or without spiral structure (early and late-type galaxies). Visual inspection was performed by the first author using the three-color SDSS images. Of the 475 central galaxies, there are 379 early-type central galaxies; of the 241 satellite galaxies, 159 are of early type. Of the 379 central galaxies, there are 217 Primary and 162 Secondary galaxies, while the 241 satellite galaxies comprise 90 and 69 Primary and Secondary galaxies (Table 1). There are 15 central and 13 satellite galaxies that we classify as ambiguous, most of which are edge-on galaxies that may be S0 or later spiral types. We exclude these ambiguous cases, although the numbers are too small to impact our conclusions. Throughout the paper we focus on the sample of early-type galaxies unless explicitly noted otherwise. 

Galaxy classification grows harder at higher redshifts. \citet{bamfordetal2009} has shown that above $z \approx 0.08$, the fraction of galaxies classified as elliptical rises unphysically as detail is lost in imaging. We could introduce a redshift-dependent fraction of fast-rotating spiral galaxies into our early-type sample if this effect is at play. The morphological bias would cause us to measure a higher early-type fraction at $z>0.08$ relative to the lower-redshift bin, caused entirely by spirals appearing as early-types. To search for this effect, we take all the galaxies with stellar masses $10^{10.75} < M_*/M_{\odot} < 10^{11}~h^{-2}$, and examine the early-type fraction with redshifts above and below $z=0.08$. There are 69 (63) objects in the low (high) redshift bin. We find consistent early-type fractions of $0.85 \pm 0.12$ and $0.80 \pm 0.13$ in the two redshift bins, suggesting that this morphology bias is not impacting our results.

Of the central, early-type galaxies, seven belong to our ancillary program. Our sample contains 30 central galaxies in halos more massive than $M_{200b}>10^{14}~h^{-1}~M_{\odot}$ and six central galaxies in halos more massive than $M_{200b}>10^{14.5}~h^{-1}~M_{\odot}$. Many central galaxies are known to have a large extended halo, sometimes known as cD galaxies \citep[e.g.,][]{morganlesh1965,schombert1984,zhaoetal2015a}; as the MaNGA sample grows it will become possible to examine trends between cD halo and $\lambda_R$.

\section{Analysis}
\label{sec:analysis}

\subsection{Kinematic Measurements}

We use the kinematic measurements provided by the MaNGA Data Analysis Pipeline (DAP; K. Westfall et al.\ in preparation). The individual spaxels are combined using Voronoi binning \citep{cappellaricopin2003} to maintain a signal-to-noise ratio of at least 10 per spectral pixel of 70~\kms. The number of combined spectra at $\sim R_e$ varies considerably from object to object depending on $R_e$, with a mean value of 12 spectra, a median value of 3 spectra, and a maximum of 200. Spaxels with individual S/N$<1$ are excluded from the binning. In calculating $\lambda_R$, we use the distance to the center of each spaxel.

The kinematics are measured using the penalized pixel-fitting code pPXF \citep{cappellariemsellem2004,cappellari2017}, with emission lines masked. Stellar templates from the MILES library  \citep{sanchezblazquezetal2006}, which cover the spectral range 3525-7500\AA\ at 2.5 \AA\ (FWHM) spectral resolution \citep{falconbarrosoetal2011}, are convolved with a Gaussian line-of-sight velocity distribution to derive the velocity and velocity dispersion of the stars (values are not corrected for instrumental resolution at this stage). An eighth-order additive polynomial is included to account for flux calibration and stellar population mismatch. 

When we extract the $\sigma$ measurements from the MaNGA catalog, we correct them for instrumental resolution using the measured $\sigma_r$, which is the unweighted average of the difference in resolution between the templates and the data as measured over all wavelengths and all spectra in the cube (K. Westfall et al.\ in prep). The MILES templates are used at their native resolution of 2.5\AA, which is higher than the MaNGA data over the full spectral range for all cubes \citep{yanetal2016b}. With this approach, the pipeline is able to reliably recover intrinsic dispersions of $\sim 35$~\kms\ or more. The MaNGA team also tried to employ the wavelength-dependent kernel convolution available within pPXF, but could not recover such low velocity dispersions with that functionality enabled. As shown by \citet{pennyetal2016}, even for intrinsic dispersions of 40 \kms, it is possible to recover the dispersion to within 10\% for high S/N spectra, while at our limiting ${\rm S/N}=10$, it is possible to recover the dispersions above the nominal limit of 70~\kms\ to within $\sim 20\%$. The DAP then supplies a single resolution correction that is the effective difference between the MILES templates and the MaNGA spectra calculated for each cube. The resolution as a function of wavelength is calculated from arcs taken before and after each exposure and then corrected using strong sky lines \citep{lawetal2016}. Tests indicate that these corrections are better than 5\% for dispersions of 70 \kms\ or higher (Westfall et al. in prep). In cases that $\sigma$ is below the spectral resolution of the instrument (a condition that does occur in the outer parts of some galaxies) we mask these values. Example maps for interesting subsets of the population are presented in Appendix A.

While in general these kinematic measurements are robust, there are some known systematic failures. There are foreground stars, which have been successfully masked by MaNGA. At lower signal-to-noise ratios, particularly in the outer regions of the galaxies, the velocity dispersion measurements can peg at the unphysical value of $1000$ km~s$^{-1}$. Finally, the superposition of two companion galaxies in the IFU field of view can lead to unphysical velocity and velocity dispersion measurements if the two components are not jointly modeled. The MPL5 catalog has identified these unphysical values; we adopt their ``DONOTUSE'' flags, as well as flagging all $\sigma >500$~\kms\ and $V>400$~\kms\ values. We only analyze objects for which at least 50\% of the fibers within $R_e$ are not flagged. Visual inspection verifies that we remove most of the clear merger cases through these cuts, and no galaxies are removed by visual inspection. Excluding these problematic cases does not lead to any systematic bias in redshift or stellar mass distributions of the galaxies. 

We perform two checks on our measurements. We explore a different analysis of the MPL4 data cubes performed by J.~Ge \citep{zhengetal2017}. The kinematics are also measured with pPXF, but the binning and masking prescriptions are different. The \lamre\ measurements agree well in general, with $\langle \lambda_{\rm MPL4} - \lambda_{\rm MPL5} \rangle = 0.02 \pm 0.09$. This provides confidence that the measurements are robust to detailed choices about masking and binning. We also compare the central $\sigma$ measurements with the original SDSS measurements that were made on different spectra using a different fitting technique. We find decent agreement, with $\langle (\sigma_{\rm MaNGA}-\sigma_{\rm SDSS})/\sigma_{\rm SDSS} \rangle = -0.05 \pm 0.1$.

\subsubsection{Radial Coverage}

Keeping only galaxies with at least $50\%$ of the fibers unmasked leaves 357 central galaxies. We have verified that no bias in stellar or halo mass is incurred when we remove the galaxies with problematic measurements. Among these 357 centrals, 272 have coverage at or beyond $R_e$, 136 have coverage at or beyond $1.5 R_e$, and 73 have coverage at or beyond $2 R_e$ (Figure \ref{fig:mass_size}; Table 1). The median effective radius of the Primary sample is $5.2$\arcsec, while the Secondary sample has a slightly smaller median of $R_e = 4.7$\arcsec. 

Turning to the satellites, there are 159 early-type galaxies with $>50\%$ of their fibers unmasked. Of these, 137 reach $R_e$ and are spatially resolved by that point, 94 reach $1.5R_e$, and 34 galaxies reach $2 R_e$. Only 50 of the early-type satellites have $R_e > 4$\arcsec\ and radial coverage out to $1.5 R_e$. In \S 3.2.1 we will return to the issue of spatial resolution and angular momentum measurements.

\begin{figure*}
\vbox{ 
\vskip -5mm
\hskip +35mm
\includegraphics[width=0.55\textwidth]{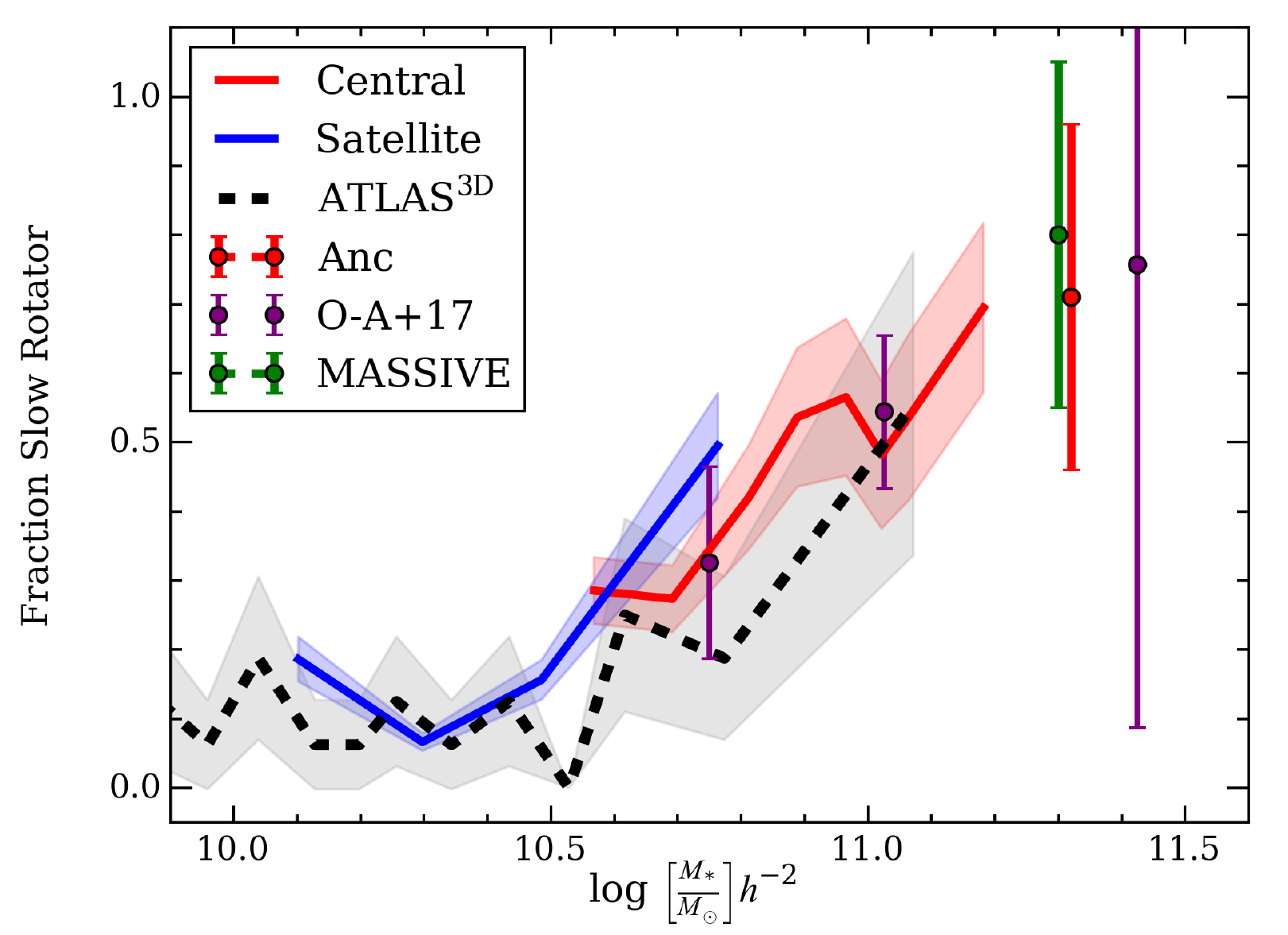}
}
\vskip -0mm
\figcaption[]{
The fraction of central (red) and satellite (blue) early-type galaxies that are slow rotators using the \citet{emsellemetal2011} criterion scaled to $1.5 R_e$; shading indicates the $1\,\sigma$ uncertainty on the fraction. The bins are chosen to contain the same number of objects, and are plotted at the weighted mean mass of each bin. We include as a separate point the central galaxies that we added to the MaNGA Primary+Secondary sample, but this represents only five systems and so has a large error bar. There is a clear trend, comparable to that reported in previous work, of a steeply rising slow rotator fraction as a function of stellar mass. We compare directly with the ATLAS$^{\rm 3D}$ fractions (grey) as a function of their stellar population mass, which we have converted from a Salpeter to a Chabrier IMF. We place the ATLAS$^{\rm 3D}$ masses in the same $h^{-2}$ units as our masses and we find good agreement with their results. We show the results from \citet{vealeetal2017a} for the MASSIVE survey, corrected to our $h^{-2}$ units, but because their stellar masses are derived dynamically, the comparison is schematic only. We also note that 20\% of the MASSIVE galaxies are satellites. Finally, the results from the highest stellar-mass bin from \citet{olivaetal2017} are presented, which shares our IMF, shifted to $h^{-2}$ units.
\label{fig:srfrac_mstar}}
\end{figure*}

\subsection{$\lambda_R$ Measurements}

The $V$ and $\sigma$ measurements on the binned data are used to calculate $\lambda_R$ (Equation 1) in elliptical apertures, as defined from the single-component \citet{sersic1963} fit from the NSA. We adopt the ellipticity [$\epsilon = 1-{\rm (B/A)}$] and position angle from this fit because the model is corrected for seeing and thus should be the most robust measurement available of these parameters, roughly measured at the effective radius (see also \S 2.3). The $R_e$ value is the circularized half-light radius from the \sers\ fit, and in the following figures and calculations, we use the radial coordinate $R_{\rm Area}$, which is calculated as the radius of the circle that would have the equivalent area as the enclosed spaxels ($R_{\rm Area} = \sqrt{R_{e,a} \times R_{e,b}} = \sqrt{N_{\rm pixels} \times A_{\rm pixel}/\pi}$), where $A_{\rm pixel}$ is the area of a pixel \citep[e.g.,][]{cappellari2013}. Examples of $\lambda_R$ are shown in Figure \ref{fig:lamradius}. In internal comparisons between different MaNGA teams, there is good agreement between \lamre\ values calculated with different prescriptions (M. Graham et al.\ in preparation).

\subsubsection{Spatial Resolution Constraints}

It is common in the literature to report $\lambda_R$ measured at the effective radius of the galaxy: \lamre. The typical galaxy in our sample has $R_e \approx 5$\arcsec. Assuming typical seeing of FWHM$ \approx 2$\arcsec, there are only four to five resolution elements across a galaxy, meaning that the $\lambda_R$ measurements are not well resolved at $R_e$. We have performed simulations (Appendix B) using the most resolved cubes in the MaNGA sample. Objects with low \lamre$<0.2$ can be highly biased towards lower \lamre\ values, by $\sim 40\%$. However, this bias can be mitigated by measuring $\lambda_R$ at larger radius. As demonstrated quantitatively in Appendix B, a decent compromise uses the outer value of $\lambda_R$, measured in the outer 10\% of the profile, \lamout. In practice, because of the typical radial extent of our data, \lamout\ matches $\lambda$($1.5 R_e$) with no bias and a scatter of $\sim 20\%$ (see Appendix B). Therefore, in what follows we will report values of \lamout, but our results do not change on average if we use the smaller sub-sample with $\lambda$($1.5 R_e$) directly available.

There is another challenging regime, those galaxies with incomplete coverage at large radius. In some cases, the $\lambda_R$ curves may not reach an asymptotic value. To quantify how often this occurs, we extrapolate each profile to $2 R_e$ using a slope fitted to the outer 20\% of the $\lambda_R$ curve. We then ask whether the fast/slow designation would change at large radius, taking into account the error in $\epsilon$ as well as in the extrapolation.  We find that 10\% of the galaxies would change designation, with these galaxies having a very similar mass distribution to the overall sample.  Thus, while limited spatial coverage is problematic, it should not change our basic results. 

\subsubsection{\lamout\ vs. $\epsilon$}

The distribution of \lamout\ as a function of $\epsilon$ is presented in Figure \ref{fig:lam_eps}. In addition to the galaxies, we show an analytic model of an edge-on galaxy in which anisotropy is proportional to ellipticity \citep[e.g.,][]{cappellarietal2007}. The majority of the fast rotators lie above this magenta line as expected from prior work \citep{emsellemetal2011}. Galaxies scatter above the magenta line because of inclination and internal variations in ellipticity. The ATLAS$^{\rm 3D}$ survey also defines an empirical division between ``slow'' and ``fast'' rotators. We will adopt a similar prescription to separate the two (black line), but in \citet{emsellemetal2011} the division between slow and fast rotator is defined at $r=R_e$. An empirical value of \lamre/$\sqrt{\epsilon}=0.31$ best divides the populations. However, we are not adopting $R_e$ as our aperture. We determine the revised value of \lamre/$\sqrt{\epsilon}$ empirically using our data. By comparing \lamout\ with \lamre, we find that the former is $14\%$ larger than the latter. Therefore, we scale the division between slow and fast rotators to \lamout/$\sqrt{\epsilon}=0.35$. A number of different divisions into fast and slow rotator have been proposed in the literature \citep[see also ][]{lauer2012,cappellari2016}. If we were to adopt the Cappellari (2016) definition instead, only $<10\%$ of galaxies would change designation.

We draw attention to the lower-right region of the figure, galaxies that apparently have relatively high $\epsilon > 0.4$ and low \lamout$<0.2$. \citet{emsellemetal2011} do not have galaxies that populate this region \citep[see also][]{cappellari2016}. Some of these outliers appear round visually, suggesting that they have poorly measured $\epsilon$ values. A related issue is that we utilize an effective $\epsilon$ value, but in general these massive galaxies grow more flattened at larger radius \citep[e.g.,][]{huangetal2013,ohetal2017}, which may contribute to the scatter. However, the majority of these galaxies are quite elongated and display low levels of rotation along their major axis, with a high central dispersion (see example maps in Appendix A). Objects in this region may be galaxies with high angular momentum that masquerade as slow rotators. In particular, galaxies known as ``double $\sigma$'' galaxies have two well-separated peaks in their $\sigma$ distributions, and have been shown to have counter-rotating disks \citep[see][and references therein]{krajnovicetal2011}. 

We visually examine the low-\lamout, high-$\epsilon$ outliers and show a possible candidate for a double-$\sigma$ galaxy in Appendix A, although our ability to distinguish such features is limited by spatial resolution \citep[see also a first sample of counter-rotating gas disks picked out from MaNGA;][]{jinetal2016}. Otherwise, there are no obvious differences between these galaxies and those with low \lamout\ but correspondingly low $\epsilon$. We conclude that double-sigma galaxies are unlikely to dominate this outlier population, and since they constitute such a small fraction of the sample, we do not remove them from consideration in what follows.

\subsubsection{Notes on the Ancillary Galaxies}

There are seven massive central galaxies in this work that were added as part of our ongoing Ancillary Program (\S 2.2). These galaxies preferentially live in the richest environments within the MaNGA footprint by design. As a result, most of them contain companions within the MaNGA IFS footprint. In two of the seven galaxies, more than 50\% of the spaxels are masked due to contamination from this substructure.  We are currently working to jointly model the kinematics from all the different substructures to build clean kinematic maps for this sample. Our work on decomposing galaxies into distinct components follows similar analysis by \citet{taboretal2017}. In the meantime, we include only five of the seven ancillary galaxies in our analysis (Table 1). 

\section{Angular Momentum Content as a Function of Stellar and Halo Mass}
\label{sec:results}

\subsection{Slow Rotator Fraction as a Function of Stellar Mass}

With \lamout\ and slow/fast rotator determinations in hand, we investigate trends in the standard slow rotator fraction using the Emsellem et al.\ definition (Figure \ref{fig:srfrac_mstar}; Table 2). Our stellar mass coverage extends only to $M_*>10^{10.5}~h^{-2}~M_{\odot}$ for the central galaxies due to the halo mass cut, while our satellite sample extends a bit lower and is limited by spatial resolution. 

\begin{figure*}
\hskip 10mm
\vbox{
\vbox{ 
\vskip -5mm
\includegraphics[width=0.85\textwidth]{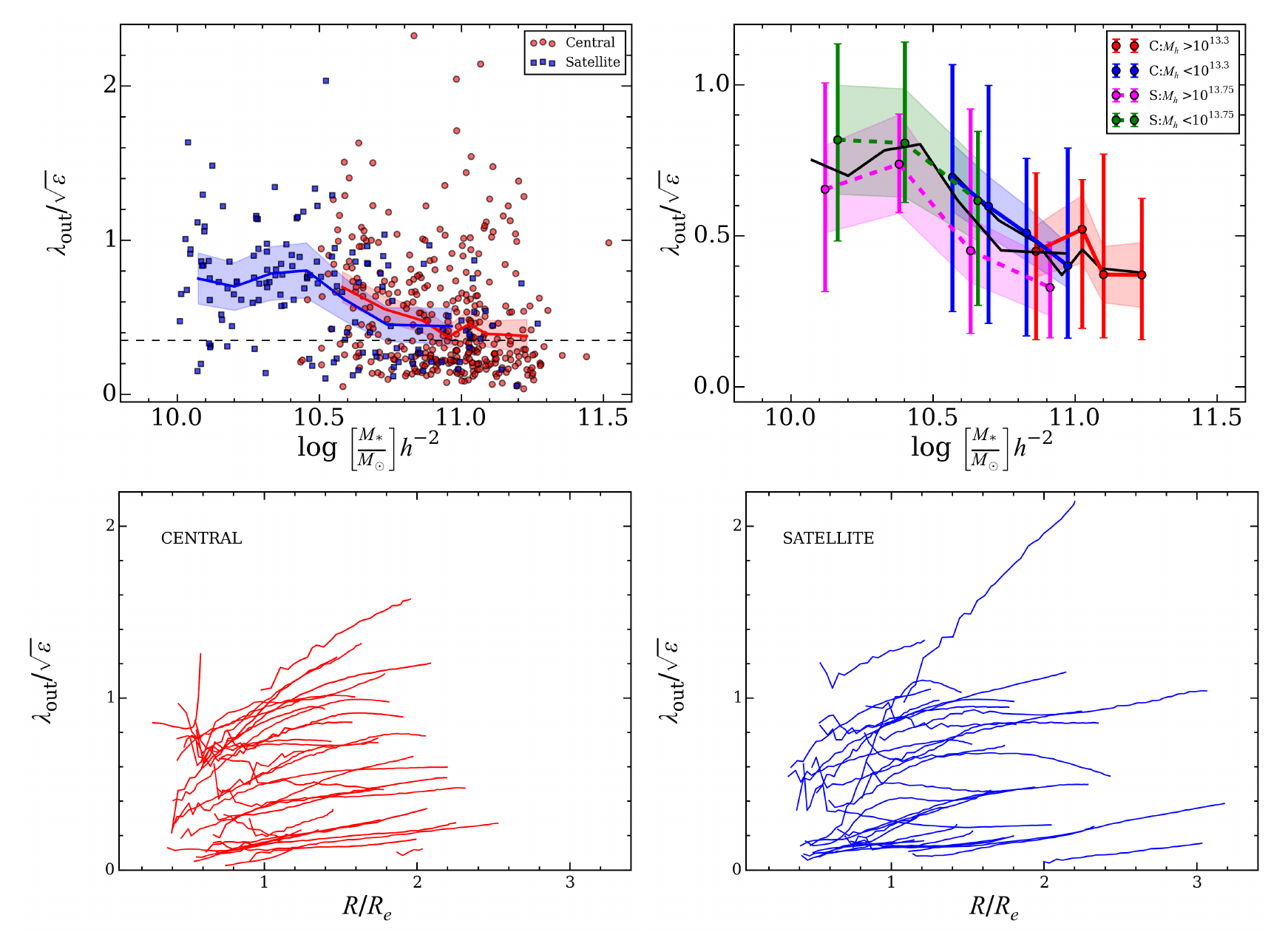}
}}
\vskip -0mm
\figcaption[]{
{\it Top Left}: The distribution of \lamout$/\sqrt{\epsilon}$, where a value of $0.35$ ({\it dashed black line}) marks the division between slow and fast rotators. Data are binned to contain equal numbers of points, and bins are plotted at the weighted mean value of $M_*$. The shaded regions denote the weighted mean and error. In the region of overlap in $M_*$ between satellite and centrals, the central galaxies appear to have a slightly higher median \lamout\ and a tail towards higher \lamout. The sharp decline in central galaxies at $M_*<10^{10.5}~h^{-2}~M_{\odot}$ comes from our halo mass cut. {\it Top Right}: The same as top left, but now also in two bins of halo mass. Shaded regions represent error in the mean, while error bars show the variance in the points. The centrals are divided at their median halo mass $M_{200b} = 10^{13.3} \, h^{-1} \, M_{\odot}$, while the satellites are divided roughly at their median halo mass, which has a higher value of $M_{200b} = 10^{13.8} \, h^{-1} \, M_{\odot}$. 
{\it Bottom Row}: Radial $\lambda_{\mathrm{out}}/\sqrt{\epsilon}$ profiles as a function of $R_{\mathrm{Area}}/R_e$. The two panels display a representative sample of satellites ({\it right}) and a mass and redshift-matched central sample ({\it left}). There is not a large difference between the two sets of profiles.
\label{fig:mstarlam}}
\end{figure*}

\hskip +10mm
\vbox{
\vbox{ 
\vskip -1mm
\includegraphics[width=0.3\textwidth]{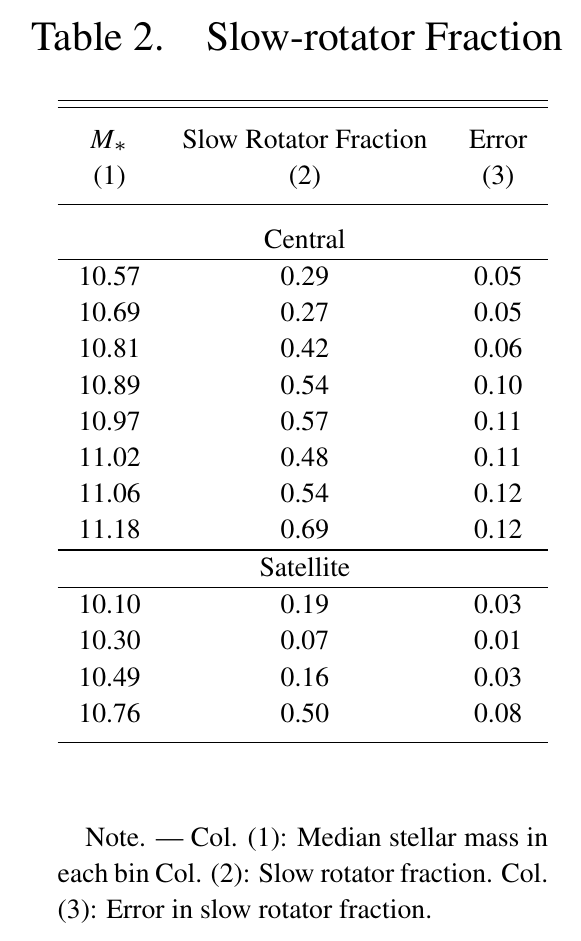}
}}
\vskip 5mm

\noindent
Also, this figure includes only the early-type galaxies, but there are likely to be some biases in these by-eye determinations \citep{bamfordetal2009}. As a sanity check, we recalculate the slow-rotator fraction with spiral galaxies included. The slow-rotator fraction changes by $\sim 50\%$ at the lowest stellar masses, but is unaffected for $M_*>10^{11}~h^{-2}~M_{\odot}$. Our conclusions are unchanged if we include the spiral galaxies in the sample.

Consistent with previous work, we see a steep increase in slow rotator fraction with stellar mass. At $M_* < 3 \times 10^{10}~h^{-2}~M_{\odot}$, galaxies are overwhelmingly fast rotators, with only $10-15\%$ slow-rotator fraction. Galaxies with stellar mass $M_* > 10^{11}~h^{-2}~M_{\odot}$ are mostly slowly rotating, and this mass scale is consistent with the quoted transition mass from \citet{cappellari2013} once our $h^{-2}$ and IMF scales are matched.

Our results are in good agreement with the literature. After correcting the ATLAS$^{\rm 3D}$ masses to be in $h^{-2}$ units and shifting them to match our assumed Chabrier IMF from Salpeter \citep[assuming a 0.2 dex decrease in mass; e.g.,][]{conroy2013} there is good agreement \citep{emsellemetal2011}. At the highest masses, our ancillary sample results agree well with the results from both \citet{vealeetal2017a} and \citet{olivaetal2017}, as well as a number of studies of individual massive clusters \citep[][]{deugenioetal2013,scottetal2014,fogartyetal2014}. Our results are also in qualitative agreement with the study by \citep{pasqualietal2007} based on the mass and luminosity dependence of the disky vs. boxy fraction of early-type galaxies, if one associates disky (boxy) galaxies with fast (slow) rotators, as suggested by the seminal work of \citet{benderetal1989} and \citet{benderetal1989}.

\subsection{Central vs.\ Satellite}

$M_*$ clearly correlates strongly with $\lambda_R$. We now address whether there is an additional dependence on large-scale environment. One approach to evaluate the impact of environment is to compare central and satellite galaxies at fixed stellar mass. 

Figure \ref{fig:srfrac_mstar} displays a trend whereby satellite galaxies have a slightly higher slow-rotator fraction than the central galaxies at fixed stellar mass. 

\vbox{ 
\vskip +0mm
\hskip +15mm
\includegraphics[width=0.25\textwidth]{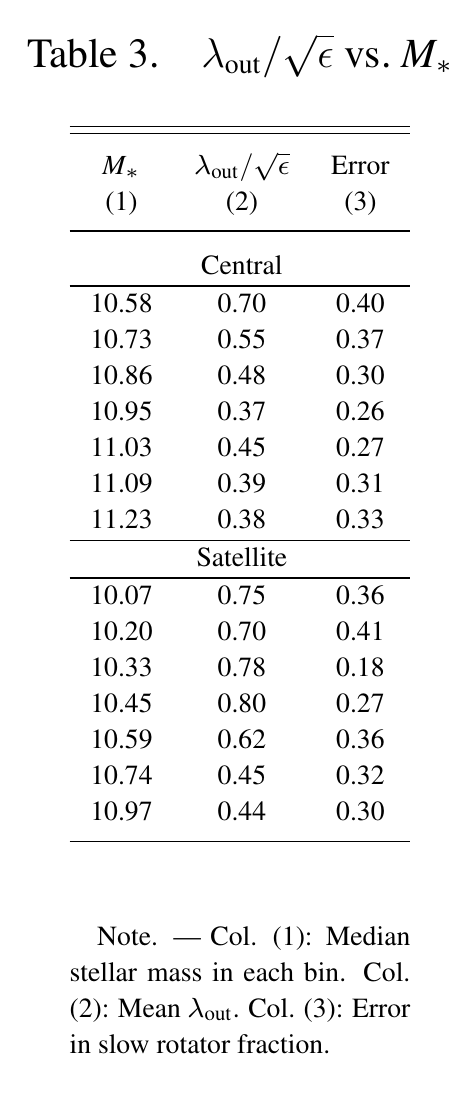}
}

\noindent
This behavior is mirrored in Figure \ref{fig:mstarlam} ({\it top left}; Table 3), where there is a difference in the median \lamout/$\sqrt{\epsilon}$, and also possibly a larger scatter towards high \lamout\ in the central relative to the satellite population. One possible driver of this difference could be spatial resolution. However, because of the design of the MaNGA survey, within the mass range of $M_* = 10^{10.5}-10^{11}~h^{-2}~M_{\odot}$, the central and satellite galaxy samples have similar median apparent sizes of 5\arcsec\ and comparable median redshifts of $\langle z \rangle = 0.065$, suggesting that spatial resolution is not obviously to blame for the systematic difference between the two populations. 

Given that the central sample is much larger than the satellite sample, and given that the detailed mass distributions do not match between the two samples, we perform an additional test to compare the satellite and central galaxies. We focus on the mass range where the two populations overlap: $M_*=10^{10.5}-10^{11}~h^{-2}~M_{\odot}$ (203 central and 54 satellite galaxies). The observed mass distribution of the satellite sample is sharply falling towards the higher mass end of this bin, while the reverse is true for the central galaxy mass distribution. We thus assign weights to the central galaxies to force the mass distribution between the two populations to match. We then build the weighted distribution in \lamout/$\sqrt{\epsilon}$ for both the central and satellite galaxies (Figure \ref{fig:massmatch}) and compare the two distributions using an Anderson-Darling test \citep[e.g.,][]{babufeigelson2006}. There is a probability of $P=33\%$ that the two samples are drawn from the same underlying distribution. There is no compelling evidence for a difference between the central and satellite galaxies in their distributions of \lamout. However, we caution that some subtle differences may still exist due to the different radial coverage between the central and satellite samples seen in Figure \ref{fig:mass_size}. Furthermore, central galaxy samples are not pure, with the level of contamination depending on halo mass \citep{skibbaetal2011}. In \S 4.5 we revisit the impact of contamination in the group catalog using the mock catalogs from \citet{campbelletal2015}.

\vbox{ 
\vskip +1mm
\hskip -5mm
\includegraphics[width=0.45\textwidth]{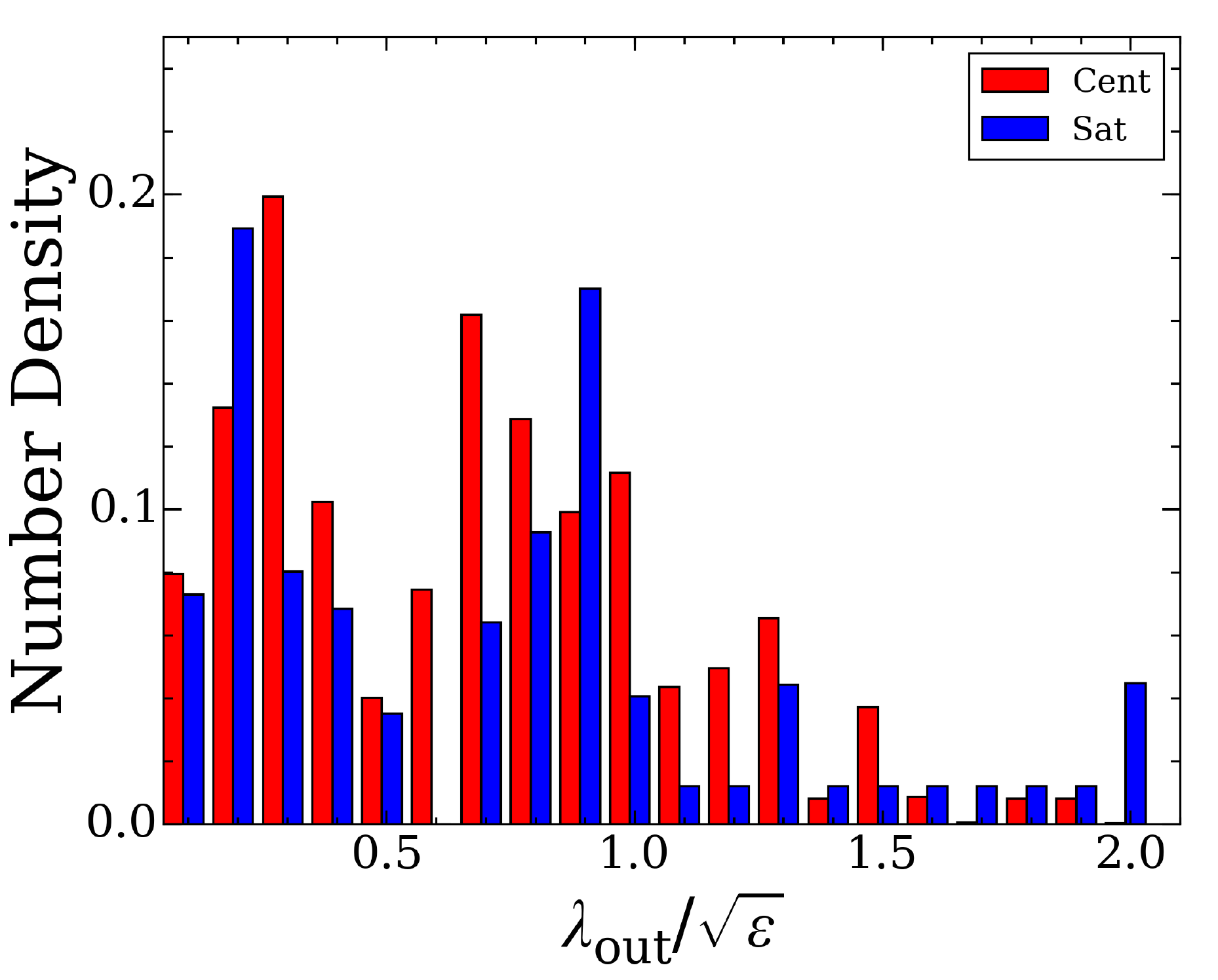}
}
\vskip -0mm
\figcaption[]{
Distribution of \lamout/$\sqrt{\epsilon}$ in the mass range $M_*=10^{10.5}-10^{11}~h^{-2}~M_{\odot}$. We apply the MaNGA weights to both samples and also re-weight the central galaxies to match the mass distribution of the satellite galaxies. When the mass distributions are carefully matched, the difference in \lamout/$\sqrt{\epsilon}$ is not significant, with an Anderson-Darling test returning a probability $P=33\%$ that the two distributions match.
\label{fig:massmatch}}
\vskip 5mm

We conclude that the central and satellite galaxies have statistically consistent distributions in \lamout/$\sqrt{\epsilon}$ when their mass distributions are carefully matched. We do not detect any significant difference between the central and satellite galaxies in their slow rotator fraction.

\subsection{Dependence on Halo Mass}

As an additional probe of the large-scale environment, we attempt to disentangle the stellar from the halo mass dependence (Figure \ref{fig:mstarlam}; {\it upper-right}). We treat the central and satellite populations separately. We divide each into two groups based on their host halo mass, and examine the weighted mean \lamout/$\sqrt{\epsilon}$ of that subpopulation. We divide the central galaxies with a halo mass above and below $10^{13.3}~h^{-2}~M_{\odot}$, which is the median halo mass. There are 172 (185) central galaxies in the higher (lower)-mass halo bin. The satellites are divided at their median halo mass of $10^{13.8}~h^{-2}~M_{\odot}$, There are 83 (76) satellites in the higher (lower)-mass bin respectively. 

We re-weight the distributions such that the stellar mass distributions match, over the mass range of overlap ($M_*=10^{10.8}-10^{11.1}~h^{-2}~M_{\odot}$) for the $\sim 100$ galaxies in each mass-limited sample. Although this comparison is limited to a narrow range in stellar mass, the distributions of central and satellite $\lambda/\sqrt{\epsilon}$ are consistent with each other, with an Anderson-Darling test returning a $P=30\%$ chance that the two samples were drawn from the same distribution. Similar results are found for the satellite galaxies. This group is divided at a higher halo mass of $M_{200b}=10^{13.8}~h^{-1}~M_{\odot}$. In a mass range of $M_*=10^{10.0}-10^{10.7}~h^{-2}~M_{\odot}$ ($\sim 60$ galaxies in each bin) we see that satellites in lower-mass halos tend to have a higher \lamout/$\sqrt{\epsilon}$. After forcing the mass distributions to match, we find only a marginally significant difference between the high and low halo masses, with an Anderson-Darling test returning a probability $P=4\%$ that the two samples are drawn from the same distribution. A larger sample is needed to investigate whether there is a real difference in the satellite population as a function of halo mass, but in the centrals our finding of no halo mass dependence is consistent with prior work \citep{vealeetal2017b,broughetal2017}. Finally, we check for a correlation between \lamout/$\sqrt{\epsilon}$ and the magnitude difference between the central galaxy and the next brightest galaxy, and find no correlation.

In a companion paper \citep{greeneetal2017}, we further compare the fast and slow rotator fractions as a function of local overdensity to compare with the recent literature \citep[e.g.,][]{cappellari2016,vealeetal2017b,broughetal2017}. 

\subsection{Galaxy Properties of Slow vs Fast Rotators}

While the majority of galaxies at high stellar mass are slowly rotating, there is a tail of fast-rotating galaxies even in the highest stellar mass bin \citep[see also][]{jimmyetal2013}. Maps of massive fast and slow rotators are shown in Appendix A. Nearby small companions add some contamination to this class of objects (at the $\sim 15\%$ level) but the majority are single objects with real rotation. We now investigate whether  properties of the fastest and slowest rotating galaxies differ in any other interesting ways.

We select the 32 fast-rotating central galaxies with \lamout$ > 0.3$ and $M_* > 10^{11}~h^{-2}~M_{\odot}$ and compare with the 43 galaxies of the same stellar mass that have \lamout$ < 0.1$. The two samples have similar median galaxy sizes of $\langle R_{\rm petro} \rangle = $16 and 15 kpc, respectively, and an Anderson-Darling test shows that they have indistinguishable distributions in size. They are at similar median redshift of $\langle z \rangle = 0.1$, and have similar flattening of $\langle \epsilon \rangle = 0.3$ and $0.2$, respectively, again with distributions consistent with arising from the same distribution. The galaxies have similar median color, $\langle g-r\rangle =0.94$~mag. The velocity dispersions in the fast rotators are lower (mean of 230 and 260~km~s$^{-1}$ respectively; $P=6 \times 10^{-4}$ that the distributions are the same). This difference is not surprising since \lamout\ depends inversely on $\sigma$.

The one independent difference between the two samples appears to be in their emission line properties. Focusing on the H$\alpha$ equivalent width within 3\arcsec, and including only systems where the emission line within that radius is measured with $>3 \sigma$ significance, there is a measurable difference in the equivalent width distribution ($P=0.003$ of belonging to the same distribution), with 47\% of the fast rotators having H$\alpha$ EW$>1$\AA, while only 16\% of the slow rotators have H$\alpha$ EW$>1$\AA\ (Fig.\ \ref{fig:massmatch}). This difference in emission-line properties suggests that the fast rotators do typically have higher gas content, as we might expect for galaxies with a disk component, perhaps associated with the event that increased their spin (see \S \ref{sec:theory} below). Unfortunately, the S/N ratios of the other strong lines are not high enough to examine the sources of photoionization (e.g., star formation, active nuclei) in these objects without more work. We perform a similar test with the satellite galaxies, but to construct a decent sample we must shift the mass limit down to log ($M_*) = 10.5~h^{-2}~M_{\odot}$. We see similar trends, although the difference in emission properties is less significant for the satellites. 

In the future, it will be interesting to investigate additional galaxy properties, such as luminosity-weighted mean age, which seems to correlate with the outer ellipticity of halos \citep[][]{ohetal2017}, disky or boxy isophotes \citep[e.g.,][]{benderetal1989}, stellar population gradients \citep[e.g.,][]{greeneetal2015,goddardetal2017}, or even cD envelope fraction, as \citet[][]{zhaoetal2015b} have argued that the cD envelope is an alternate tracer of merger history.

\subsection{Impact of Imperfect Group Catalogs}
\label{sec:mocks}

Group finders cannot perfectly identify dark matter halos. For our purposes, there are two main concerns. There is scatter in the assignment of halo mass to groups, leading to uncertainties in the halo masses. Also, there are errors in the identification of central galaxies \citep[e.g.,][]{skibbaetal2011,linetal2016,langeetal2017}. To explore how these two errors impact our results, we employ a mock catalog built by \citet{campbelletal2015}. Campbell et al.\ use N-body simulations to create a ``true'' galaxy catalog matched to the SDSS with perfect knowledge of the groups, halo masses, and central galaxies. They then run the Y07 algorithm on their mock galaxy catalog to create a mock Y07 group catalog. This catalog contains the same scatter in halo mass and in central designation as Y07, being constructed in the same manner.  

\vbox{ 
\vskip +1mm
\hskip -5mm
\includegraphics[width=0.45\textwidth]{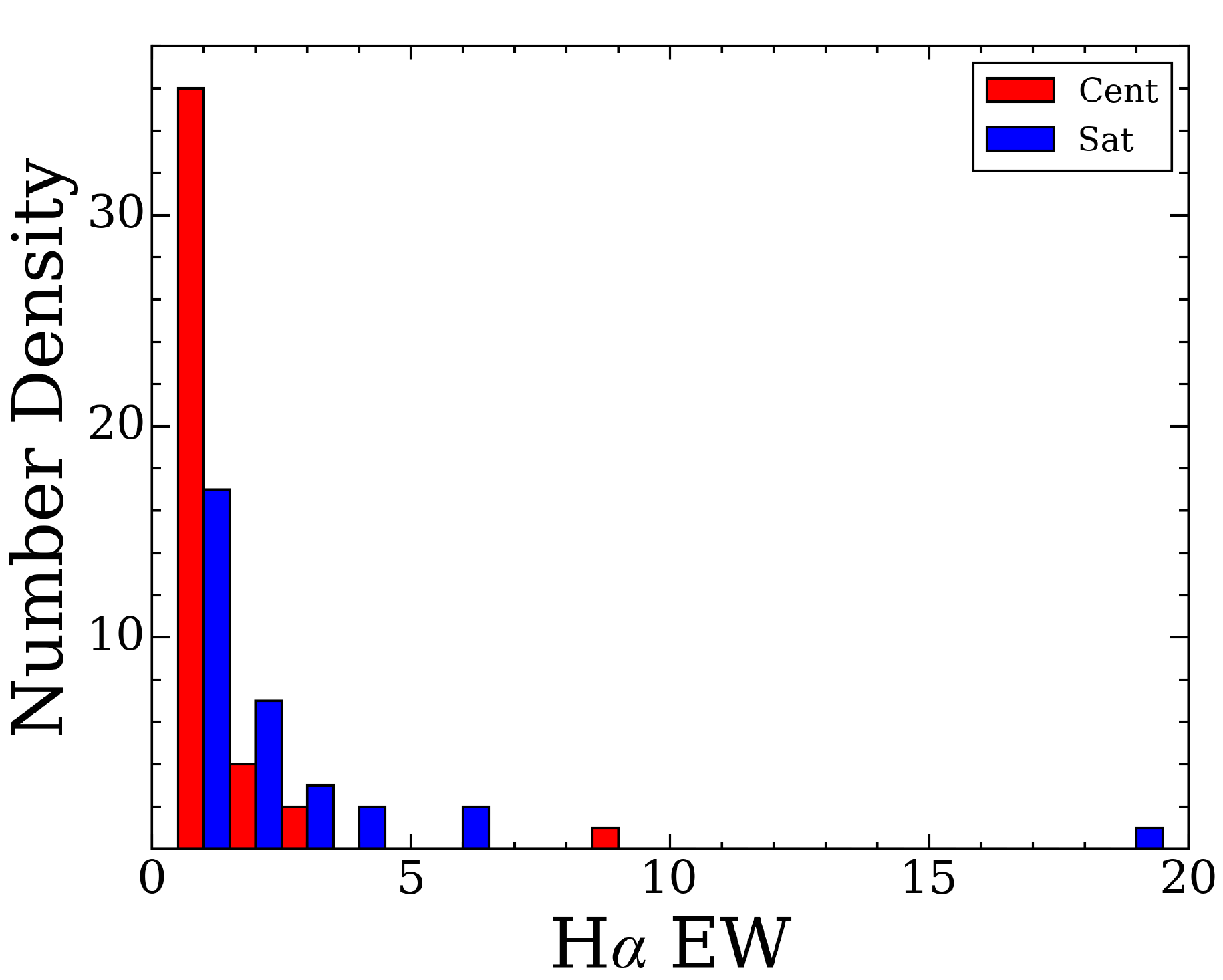}
}
\vskip -0mm
\figcaption[]{
Distribution of H$\alpha$ equivalent widths measured at the galaxy center for central galaxies with masses $M_*>10^{11}~h^{-2}~M_{\odot}$ and \lamout$>0.3$ (blue) or \lamout$<0.1$ (red). The fast rotators have a broader distribution towards high equivalent widths.
\label{fig:massmatch}}
\vskip 5mm

We assign every mock galaxy a value of \lamout/$\sqrt{\epsilon}$ based on its stellar mass, with a linear dependence designed to roughly reproduce our measurements. We model the \lamout/$\sqrt{\epsilon}$ distribution as a sum of two Gaussians. The primary Gaussian (containing 80\% of the galaxies) has a narrow width of 0.2, and a central value that varies with stellar mass as \lamout/$\sqrt{\epsilon} = -0.7 {\rm log} M_* - 0.14$. The secondary Gaussian has no mass dependence and a large scatter and is added in an ad hoc way to match the scatter that we see at all masses. The center of the distribution is fixed at \lamout/$\sqrt{\epsilon}=0.8$ with a dispersion of 0.5, and this component comprises 20\% of the total. In the fiducial assignment, satellite and central galaxies are treated the same way at fixed stellar mass.

\begin{figure*}
\hskip 15mm
\vbox{ 
\includegraphics[width=0.4\textwidth]{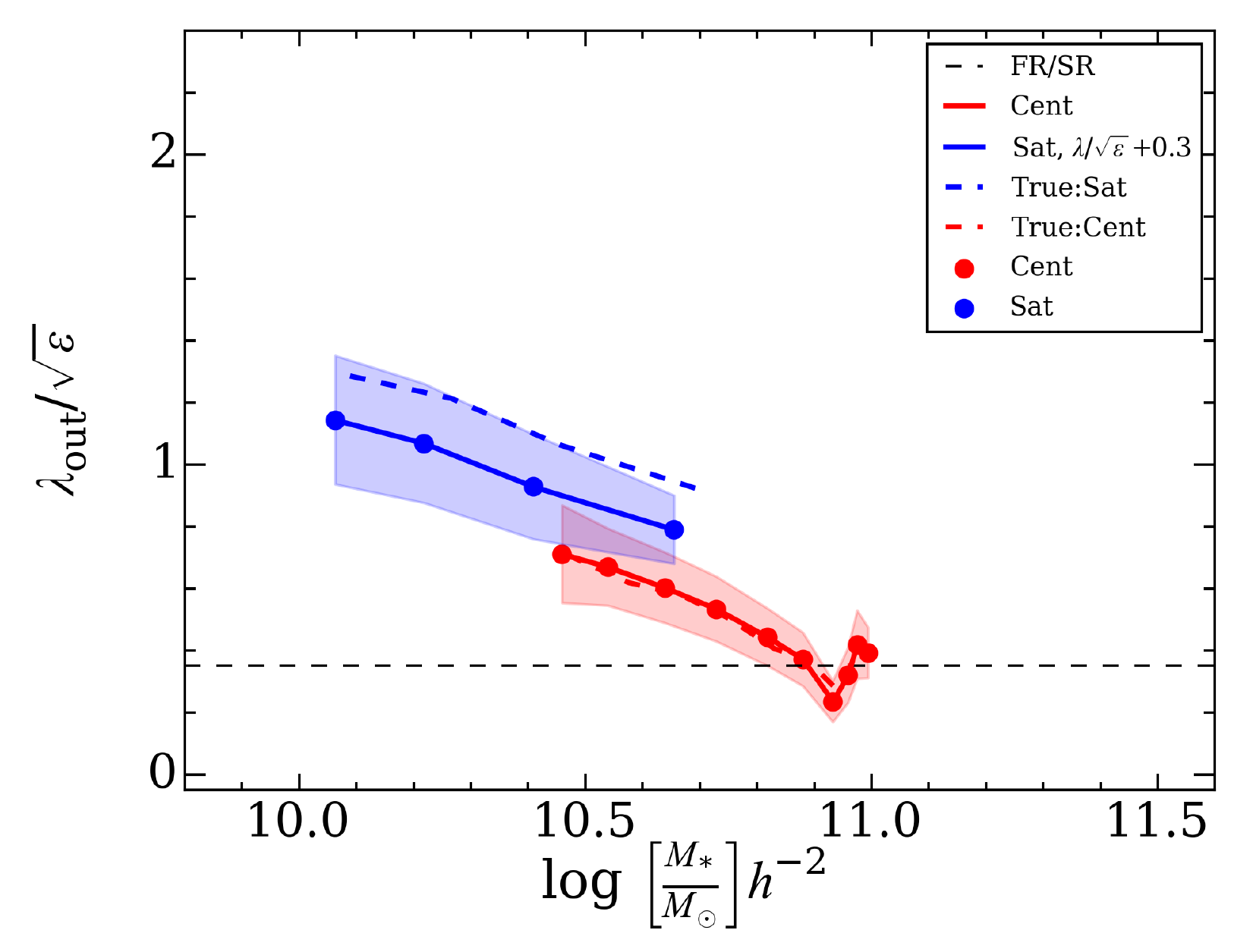}
\includegraphics[width=0.415\textwidth]{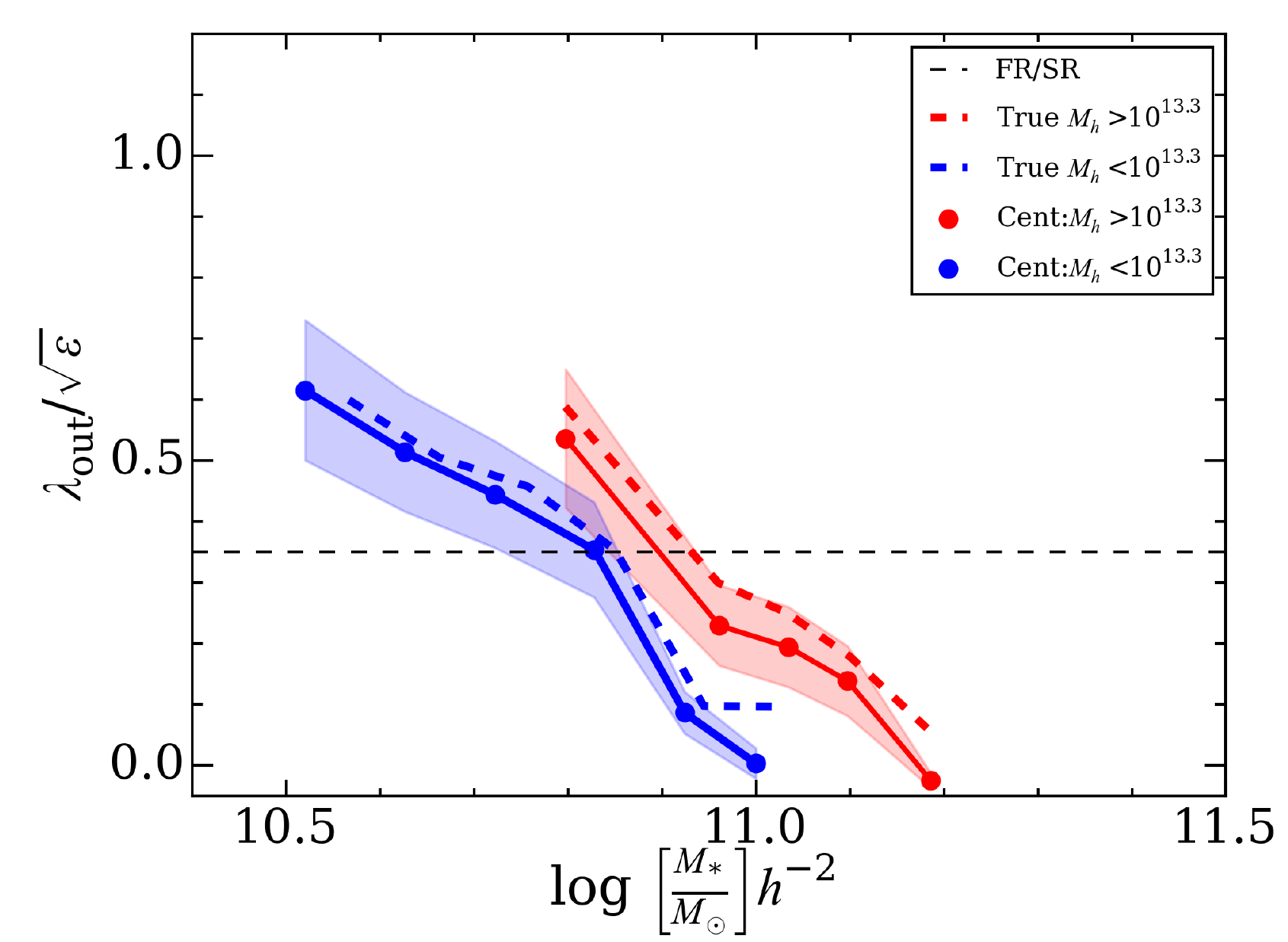}
}
\vskip -0mm
\figcaption[]{
{\it Left}: Weighted mean \lamout$/\sqrt{\epsilon}$ for the mock catalog built from 100 random draws to yield the same sample size and demographics as our real data. Central galaxies (red) and satellites (blue) are offset in their $\lambda/\sqrt{\epsilon}$ by a Gaussian random value with mean 0.35 and scatter 0.05. Binning and weighting done as above in Figure \ref{fig:mstarlam}, and the input relations are shown with the dashed line for comparison. Although errors in central/satellite determination do add a systematic bias to our output relation between stellar mass and $\lambda/\sqrt{\epsilon}$, we are sensitive to difference between the two populations.
{\it Right}: Weighted mean \lamout$/\sqrt{\epsilon}$ for the mock catalog, now just examining central galaxies in two $M_{200b}$ bins as indicated. Here we introduce a halo-mass dependent scatter to \lamout$/\sqrt{\epsilon}$, such that galaxies at fixed stellar mass with higher halo mass are given a higher value of \lamout$/\sqrt{\epsilon}$ [$\delta$ \lamout$/\sqrt{\epsilon}$ = m (log $M_{200b}$ - log $M_{200b, median}$), for slope m]. Despite systematic bias due to scatter in halo mass, this difference is detectable when the slope $m>0.3$.
\label{fig:mocks}}
\end{figure*}

First, we use these catalogs to test our sensitivity to differences between central and satellite populations. We create a suite of simulated satellite galaxies in which \lamout/$\sqrt{\epsilon}$ is boosted relative to the default values for central galaxies at a given stellar mass by $\delta$\lamout/$\sqrt{\epsilon} = 0.1, 0.2, 0.3, 0.35$ on average, with a scatter of 0.05. These differences are introduced in the true catalog, and then we ask whether we can recover this difference in the Y07 mock catalog. We then create 100 data sets with statistics matched to our true sample by selecting a mock galaxy with stellar mass within 0.05 dex and halo mass within 0.1 dex of each sample galaxy, for both central and satellite galaxies. We then calculate the dependence of \lamout/$\sqrt{\epsilon}$ and slow-rotator fraction on $M_*$ as with the real data (adopting the appropriate MaNGA weight for each mock galaxy). We find that the difference between satellite and central becomes measurable when the offset is $\delta$\lamout/$\sqrt{\epsilon} = 0.2$ and significant when the offset is $\delta$\lamout/$\sqrt{\epsilon} = 0.35$ (Figure \ref{fig:mocks}). As is apparent from the figure, there is a bias introduced by the mixing between satellites and centrals.

Second, we test whether we could uncover a secondary trend in \lamout$/\sqrt{\epsilon}$ with $M_{200b}$ at fixed stellar mass. We scatter the default values in the catalog by an amount that depends on halo mass. Specifically, we perturb the values according to: $\delta$ \lamout$/\sqrt{\epsilon}$ = $m$ (log $M_{200b}$ - log $M_{\rm 200b, median}$), for slope m, with $m=0.1, 0.2, 0.3$ and a scatter in m of 0.05. In Figure \ref{fig:mocks} we see that there is minimal bias introduced by scatter in halo masses, and the two halo-mass bins become measurably different for a slope of $>0.3$, corresponding to changes in \lamout$/\sqrt{\epsilon} \sim 0.2$ at $M^* \approx 10^{11}$~\msun. 

We see that the errors in the satellite/central designation suppress the input difference between central and satellite. The amount of suppression ($\lambda/\sqrt{\epsilon} \sim 0.2$) is of the same order as the current uncertainty in the mean value. Thus, simply increasing the total number of objects will not help uncover subtle differences between central and satellite galaxies; rather we are limited by the systematic errors in the designation of centrals and satellites by group finders. While it is possible to uncover trends in halo mass for central galaxies (test 2 above), to distinguish differences in satellite vs. central galaxies, the impact of group finding methods will have to be forward modeled via mock catalogs. 

\subsection{Linking $\lambda_R$ with formation history}
\label{sec:theory}

Several papers have used simulations to investigate the primary mechanisms that impact $\lambda_R$ and galaxy flattening ($\epsilon$) in galaxies. We compare our results with their predictions here.

Several studies \citep[e.g.,][]{khochfarburkert2005,naabetal2006,kangetal2007} have argued that the dichotomy of early-type galaxies can be explained in a scenario whereby boxy, slowly rotating ellipticals have their origin in a merger that is {\it both} major (i.e., mass ratio of progenitors close to unity) {\it and} dry (i.e., progenitors have small gas mass fractions). In particular, \citet{kangetal2007} conclude that the observed stellar mass dependence of the boxy fraction requires that slow rotators result from mergers with a progenitor mass ratio < 2 and with a combined cold gas mass fraction < 0.1. \citet[][]{lagosetal2017} also find that major merging is a primary driver of angular momentum evolution. Interestingly, \citet[][]{choiyi2017} find that the cluster galaxies in their simulations with no major merging are the ones with the most rapid decrease in $\lambda_R$, but they are not certain what physical process drives this decline. 

\citet{penoyreetal2017} examine the distribution of $\lambda_R$ with stellar mass using the Illustris simulation \citep{vogelsbergeretal2014}. They report that major mergers and gas accretion have the strongest impact on $\lambda_R$, with the former typically spinning down and the latter spinning up galaxies, although they do not yet consider black hole feedback or large-scale intrinsic alignments of galaxies as possible factors \citep[see][ for the possible importance of black hole feedback]{martizzietal2014}. They find that lower mass galaxies can be spun up by accretion of gas, while at higher mass the accretion rates are not high enough to change $\lambda_R$. They suggest that high-mass galaxies above $M_* \approx 10^{11}~M_{\odot}$ have uniformly low $\lambda_R$ because at later times mergers and accretion are unable to significantly change their angular momentum content. In contrast, lower-mass galaxies can be spun up at late times by accretion and star formation. \citet{penoyreetal2017} also find that faster-rotating galaxies are more gas rich (and more metal-enhanced) and they do not find a discernible difference between satellite and central galaxies at fixed mass. 

\citet{naabetal2014} use cosmological simulations of 44 massive galaxies to examine the relationship between merger history and angular momentum content. In addition to $\lambda_R$, they compare their simulations with observations of the flattening of the merged remnant ($\epsilon$), and higher-order moments of the line-of-sight velocity distribution parametrized with Gauss-Hermite polynomials; $h_3$ is the asymmetry parameter \citep[e.g.,][]{vandermarelfranx1993}. They identify three pathways that form fast rotators: galaxies with little merging but some gas (e.g., a faded disk galaxy), late gas-rich merging that spins up the remnant (these have anti-correlated $h_3$ and $V/\sigma$), and late dissipationless merging that spins up the remnant \citep[no anti-correlation between $h_3$ and $V/\sigma$;][]{naabburkert2001,naabetal2006}. \citet{vandesandeetal2017} have identified SAMI galaxies without an $h_3$-$V/\sigma$ anticorrelation that may indeed be the remnants of late dissipationless merging. There are also three main pathways to make slow rotators in their simulations: galaxies that form early and only experience minor merging since, galaxies with a gas-rich major merger that spins down the remnant, and galaxies with a gas-poor major merger that spins down the remnant. Apparently these last two events produce highly flattened configurations, and may correspond to our small tail of high \lamout, high $\epsilon$ galaxies. 

\section{Summary}
\label{sec:summary}

We present an unprecedented sample of 503 (379 early-type) central and 241 (159 early-type) satellite galaxies observed with the SDSS-IV MaNGA IFU survey. We leverage this sample to study the dependence of the specific stellar angular momentum $\lambda_R$ on stellar and halo mass. We define a new measure of $\lambda$, the asymptotic value \lamout, that allows us to compare cubes with different spatial resolution and spatial coverage. We investigate the slow rotator fraction along with the \lamout\ distributions as a function of stellar ($M_*$) and halo ($M_{200b}$) mass. Overall, the observed distribution of galaxies in the \lamout\ vs $\epsilon$ plane matches expectations from previous work, with most fast rotators well-described by oblate rotator models in which the anisotropy correlates with $\epsilon$. However, there is a small but interesting tail of galaxies with low \lamout\ and high $\epsilon$.

Aligned with all previous work on this topic, we find a clear and strong dependence of \lamout/$\sqrt{\epsilon}$ on stellar mass $M_*$. There is a tail of high angular momentum galaxies ($\sim 30\%$) even at the highest masses. These galaxies tend to contain more ionized gas emission, but otherwise show no other differences with the slowly-rotating systems. Central and satellite galaxies have similar slow-rotator fractions and distributions in \lamout/$\sqrt{\epsilon}$ at fixed stellar mass. There is no evidence for a residual dependence of \lamout/$\sqrt{\epsilon}$ on $M_{200b}$ for central or satellite galaxies once the mass distributions are matched. 

As the MaNGA survey continues, the number of massive central galaxies in the most massive halos will increase, producing a wider baseline for study. We will also investigate secondary trends with stellar populations and gradients therein \citep[e.g.,][]{greeneetal2015,ohetal2017,goddardetal2017}, the spatial extent of the ionized gas, and the presence or absence of an extended stellar halo.

\begin{acknowledgements}

We thank the anonymous referee for carefully reading the manuscript multiple times. We thank P. Hopkins and K. Bundy for useful conversations. We especially thank Z. Penoyre and S. Genel for providing their manuscript before submission. J.E.G. is partially supported by NSF AST-1411642. SM is supported by the Japan Society for Promotion of Science grants JP15K17600 and JP16H01089. FB acknowledges support from the Klaus Tschira Foundation, through the HITS-Yale Program in Astrophysics (HYPA). 

Funding for the Sloan Digital Sky Survey IV has been provided by
the Alfred P. Sloan Foundation, the U.S. Department of Energy Office of
Science, and the Participating Institutions. SDSS-IV acknowledges
support and resources from the Center for High-Performance Computing at
the University of Utah. The SDSS web site is www.sdss.org.

SDSS-IV is managed by the Astrophysical Research Consortium for the 
Participating Institutions of the SDSS Collaboration including the 
Brazilian Participation Group, the Carnegie Institution for Science, 
Carnegie Mellon University, the Chilean Participation Group, the French Participation Group, Harvard-Smithsonian Center for Astrophysics, 
Instituto de Astrof\'isica de Canarias, The Johns Hopkins University, 
Kavli Institute for the Physics and Mathematics of the Universe (IPMU) / 
University of Tokyo, Lawrence Berkeley National Laboratory, 
Leibniz Institut f\"ur Astrophysik Potsdam (AIP),  
Max-Planck-Institut f\"ur Astronomie (MPIA Heidelberg), 
Max-Planck-Institut f\"ur Astrophysik (MPA Garching), 
Max-Planck-Institut f\"ur Extraterrestrische Physik (MPE), 
National Astronomical Observatories of China, New Mexico State University, 
New York University, University of Notre Dame, 
Observat\'ario Nacional / MCTI, The Ohio State University, 
Pennsylvania State University, Shanghai Astronomical Observatory, 
United Kingdom Participation Group,
Universidad Nacional Aut\'onoma de M\'exico, University of Arizona, 
University of Colorado Boulder, University of Oxford, University of Portsmouth, 
University of Utah, University of Virginia, University of Washington, University of Wisconsin, 
Vanderbilt University, and Yale University.

This research made use of Marvin, a core Python package and web framework for MaNGA data, developed by Brian Cherinka, Jose Sanchez-Gallego, and Brett Andrews (MaNGA Collaboration, 2017).

\end{acknowledgements}


\newpage 

\appendix

\section{Appendix A}
\label{sec:appendixa}

Three figures displaying examples of galaxies with high stellar mass ($M_*>10^{11}~h^{-2}~M_{\odot}$) and \lamout$<0.1$ (Figure A8), galaxies in the same mass range but with \lamout$>0.4$, and finally galaxies with \lamout$<0.15$ and $\epsilon>0.45$.

\begin{figure*}
\vskip -25mm
\hskip 15mm
\vbox{
\vbox{ 
\includegraphics[width=0.75\textwidth]{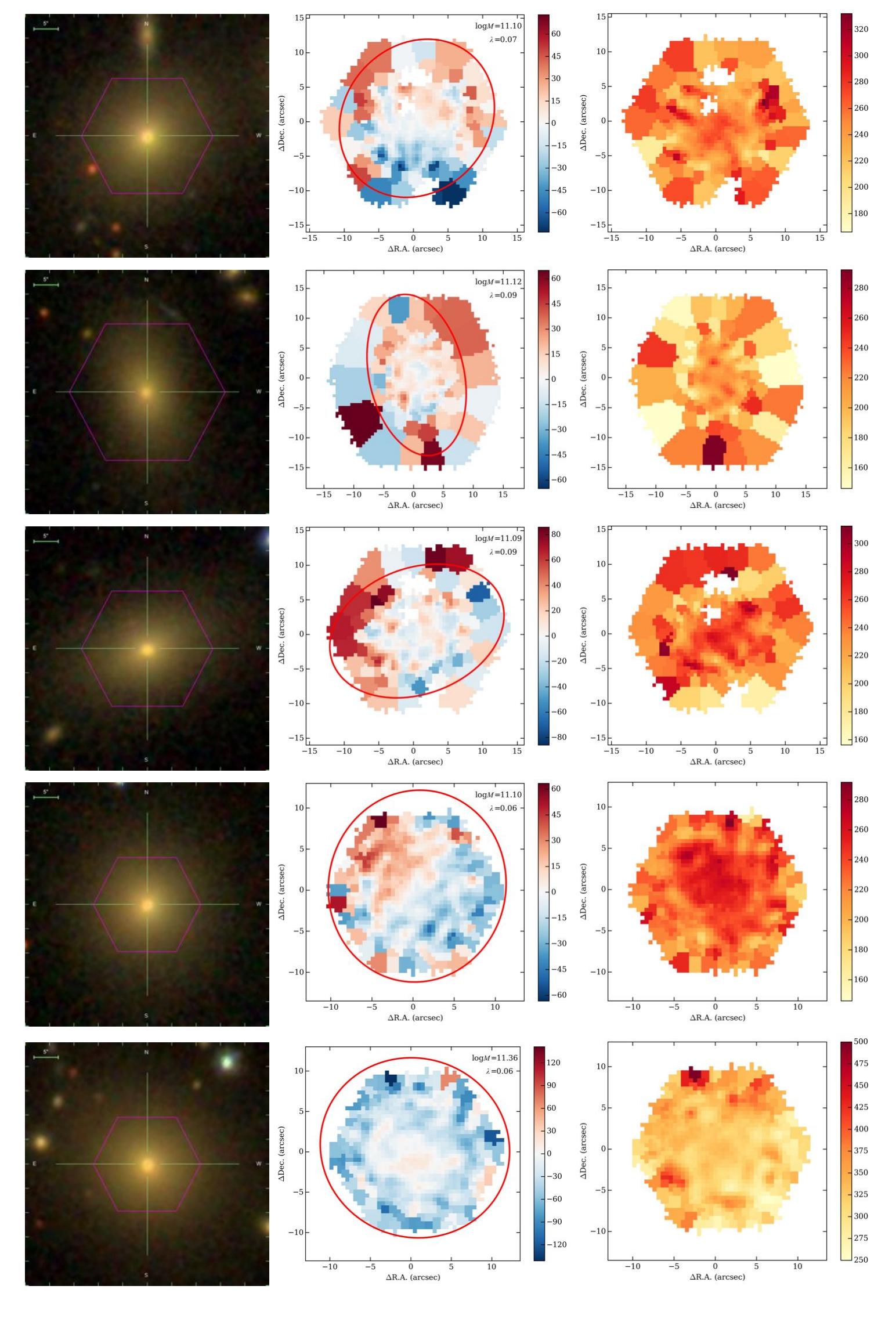}
}
}
\figcaption[]{High-mass slow rotators. For each galaxy we present the three-band ($gri$) SDSS image with the MaNGA footprint superimposed (left) and the velocity field (middle) and velocity dispersion field (right). Spaxels with $|V|>400$~km~s$^{-1}$ or $\sigma>500$~km~s$^{-1}$ are not included in our calculation of \lamout. The scale-bar represents 5\arcsec. The red ellipse superimposed on the velocity field has the effective radius, ellipticity, and PA of the galaxy fit.
\label{fig:maps-anc}}
\end{figure*}
\vskip 5mm

\begin{figure*}
\hskip 10mm
\vbox{
\vbox{ 
\vskip -15mm
\includegraphics[width=0.85\textwidth]{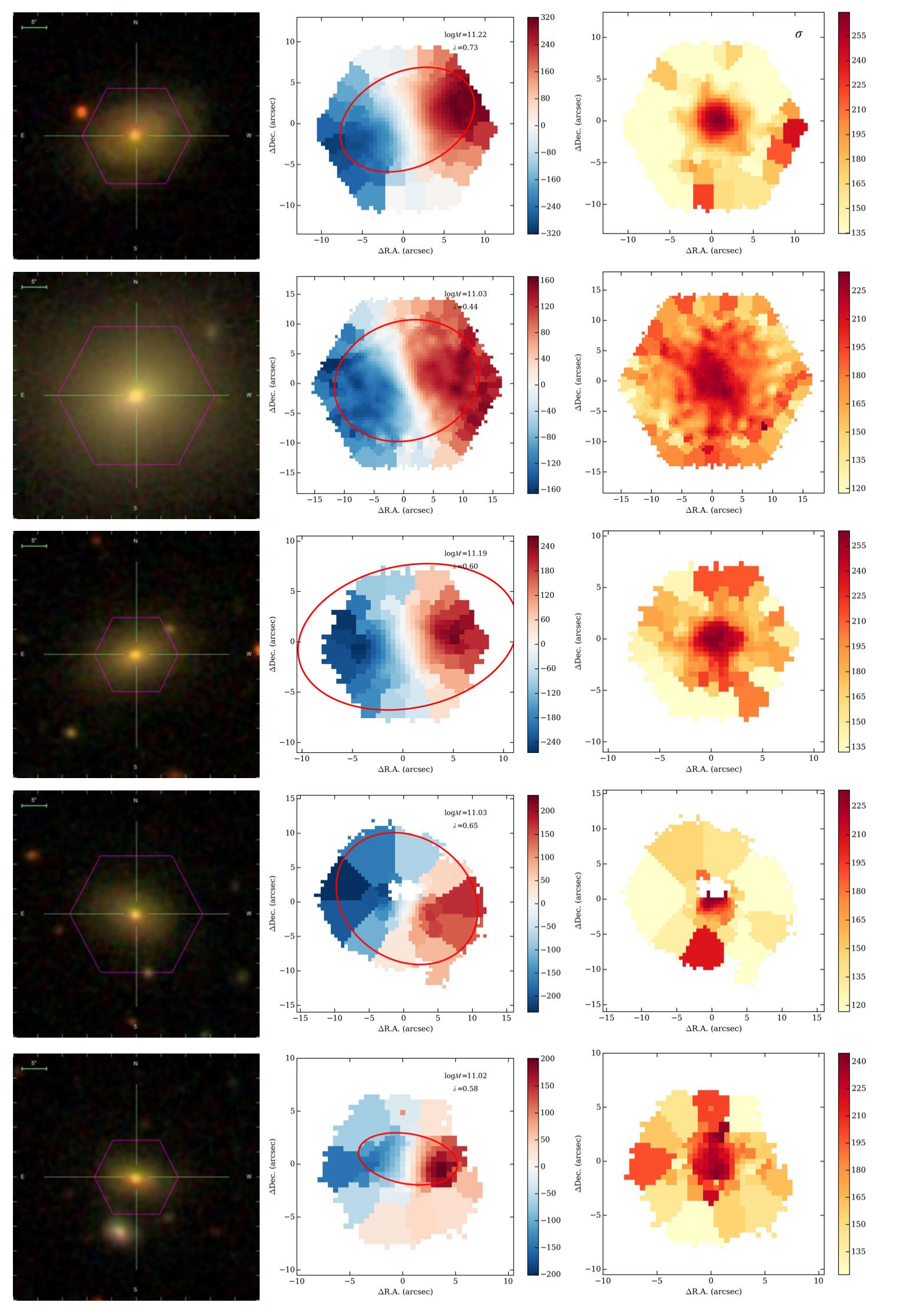}
}
}
\vskip -0mm
\figcaption[]{Examples of galaxies that have high mass ($M_*>10^{11}~h^{-2}~M_{\odot}$) and are classified as fast rotators (\lamout$>0.4$). For each galaxy we present the SDSS three-band image ($gri$; left), velocity field (middle), and velocity dispersion field (right). Spaxels with $|V|>400$~km~s$^{-1}$ or $\sigma>500$~km~s$^{-1}$ are not included in our calculation of \lamout. It is interesting to note that many of these galaxies are also quite round but do show clear rotation fields. The red ellipse superimposed on the velocity field has the effective radius, ellipticity, and PA of the galaxy fit.
\label{fig:maps-hmfr}}
\end{figure*}
\vskip 5mm

\begin{figure*}
\hskip 10mm
\vbox{
\vbox{ 
\vskip -15mm
\includegraphics[width=0.85\textwidth]{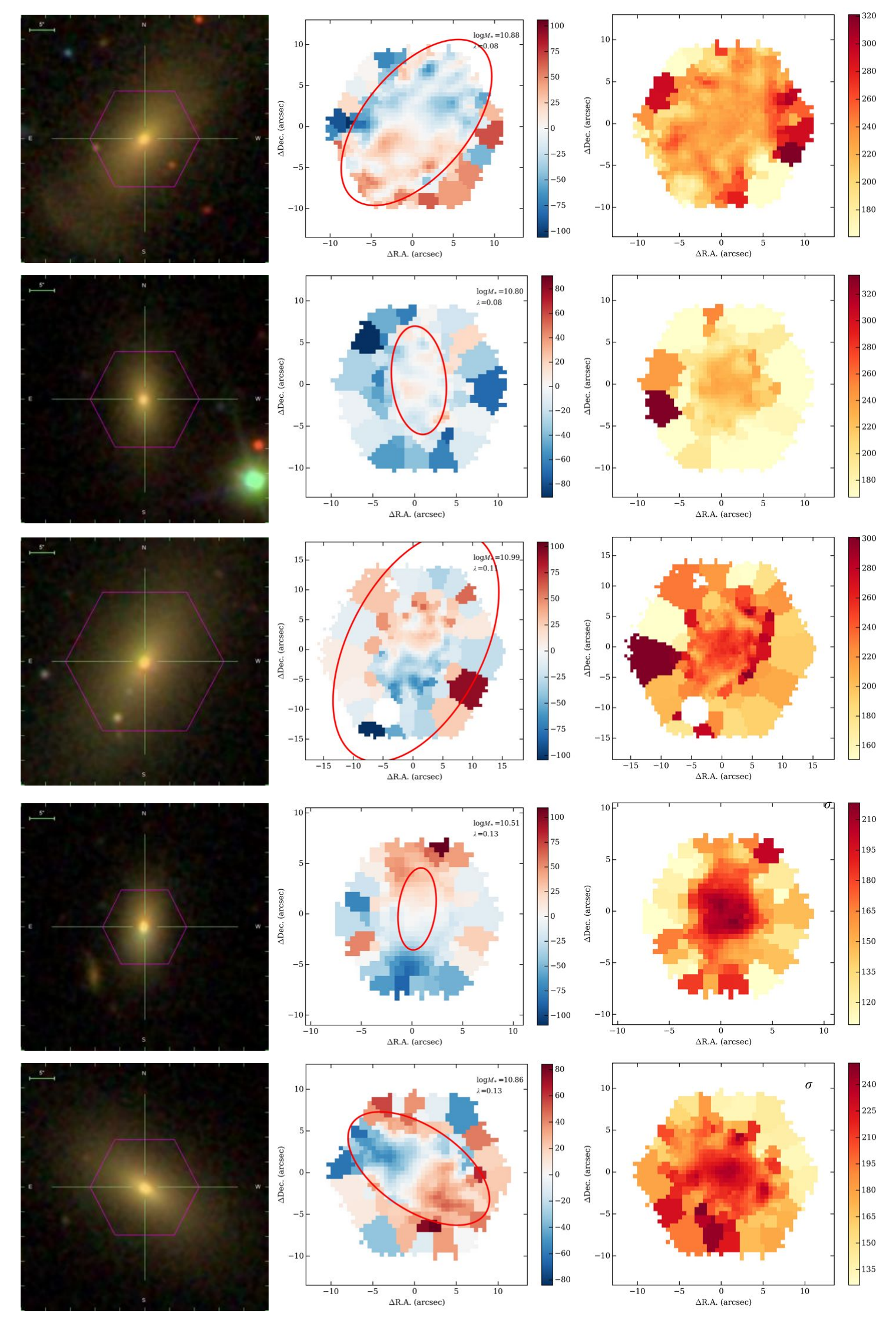}
}
}
\figcaption[]{High $\epsilon$, low $\lambda$ outliers. For each galaxy we present the three-band ($gri$) SDSS image (left), the velocity field (middle) and velocity dispersion field (right). Spaxels with $|V|>400$~km~s$^{-1}$ or $\sigma>500$~km~s$^{-1}$ are not included in our calculation of \lamout. The IFU footprint is shown on the image, along with a 5\arcsec\ scale bar. The final two galaxies in this example list are potential double-$\sigma$ galaxies. The red ellipse superimposed on the velocity field has the effective radius, ellipticity, and PA of the galaxy fit.
\label{fig:maps-anc}}
\end{figure*}
\vskip 5mm

\newpage

\section{Appendix B}

\subsection{Binning and Signal-to-Noise}

Voronoi binning is done to keep the signal-to-noise (S/N) ratio above a baseline level of 10. However, some of the outermost bins have lower S/N in practice. They also can cover large regions of the galaxy. It is natural to ask whether the $\lambda_R$ measurements are sensitive either to degraded S/N in the outermost parts of the galaxy or to irregular bin shapes. We briefly address each concern here.

\begin{figure*}
\hskip 5mm
\vbox{ 
\includegraphics[width=0.42\textwidth]{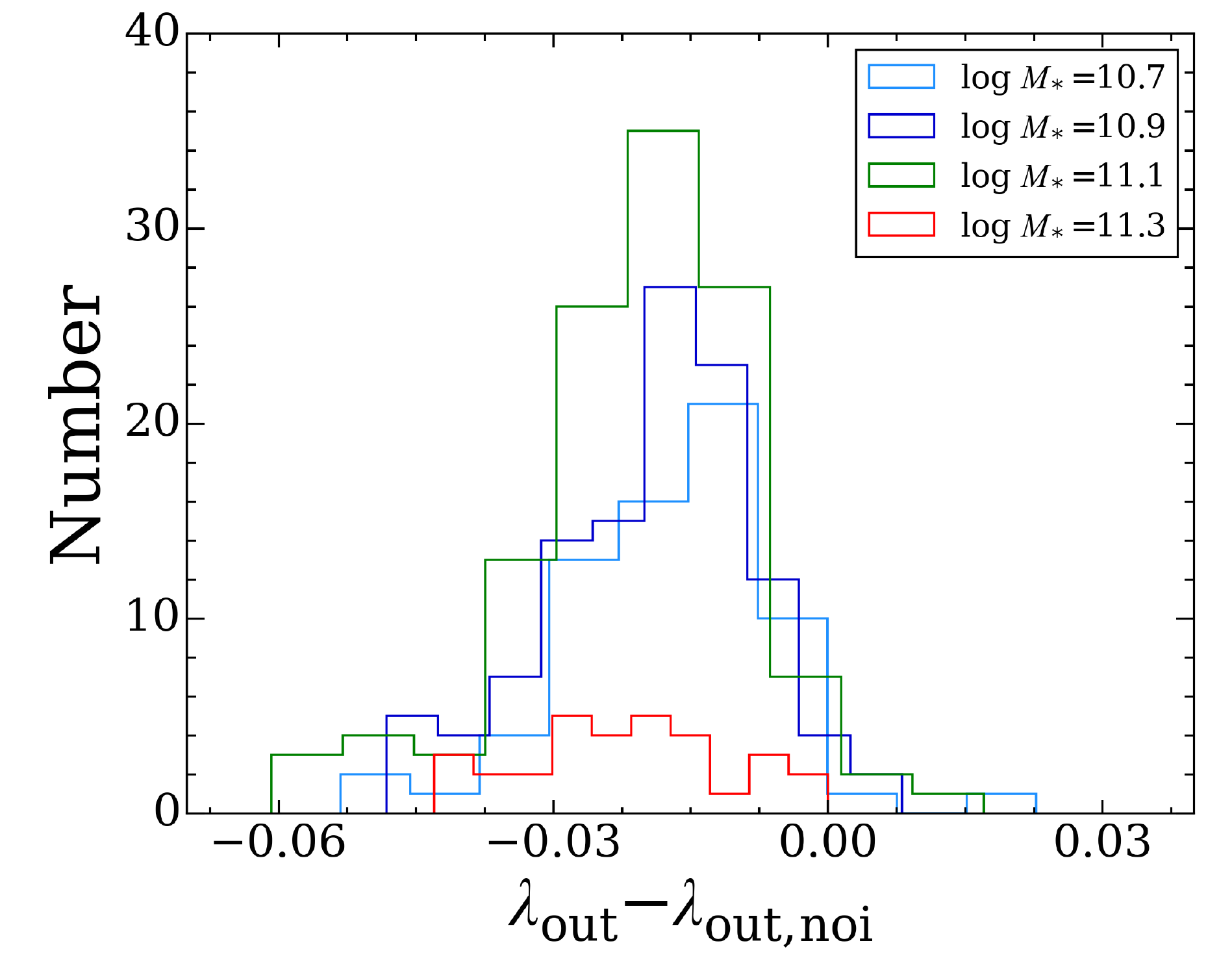}
\includegraphics[width=0.45\textwidth]{BFig0b.pdf}
}
\vskip -0mm
\figcaption[]{
{\it Left}: Histograms of the difference $\lambda_{\rm out}-\lambda_{\rm out, noi}$ as a function of stellar mass. Generally there is an offset of $-0.02 \pm 0.01$ (with the noisy measurements being a bit larger) but this difference is very small and there is no mass dependence apparent in the typical offset or the scatter.
{\it Right}: Plot of the difference $\lambda_{\rm 1.5 R_e, m} - \lambda_{\rm out}$ to show that $\lambda_{1.5 R_e}$ matches $\lambda_{\rm out}$ with 20\% scatter and no bias as a function of stellar mass.
\label{fig:deltahist}}
\end{figure*}

First, we ask whether including noisy measurements in the outer bins has a strong impact on the value of \lamout\ that we infer. We perform a simple test. We simply perturb each velocity measurement with a random Gaussian variate using twice the reported error in the measurement, and then recompute the perturbed \lamout$_{,p}$. The distribution in $\delta$\lamout$=$\lamout$-$\lamout$_{,p}$ is strongly peaked at zero, with a negative tail. That is, adding noise tends to increase \lamout\ slightly, with a median $\delta \lamout = -0.001$, with 30\% of objects having $\delta \lamout < -0.05$ and 5\% of objects having $\delta \lamout < -0.1$. We do the same test with $\sigma$ rather than $V$. This change has even less effect, presumably because $\sigma$ is an even quantity, and so the average $\sigma$ is the same everywhere at a given radius, unlike $V$. The distribution in $\delta \lamout$ is peaked at zero, with a FWHM of $0.01$. Only 8\% of systems have $\lvert \delta \lamout \rvert >0.02$ and only one object has $\lvert \delta \lamout \rvert >0.04$. Finally, we repeat the same test but introduce scatter in both $V$ and $\sigma$ at the same time. In this case the systematic bias is slightly larger, with a median $\delta \lamout = -0.02 \pm 0.01$ (Figure \ref{fig:deltahist}). Only 6\% of systems have $\lvert \delta \lamout \rvert >0.04$ and 1\% of systems have $\lvert \delta \lamout \rvert >0.05$.

The punchline is that scatter in the outer measurements does not influence \lamout\ strongly. While we experimented with excluding measurements based on large uncertainties, this test suggests that it is worth keeping all the measurements, even those that are noisy. If we exclude lower S/N measurements, we simply restrict the radial range of the data unnecessarily.

Second, we do the following test to see how the large and irregular bin sizes impact \lamout. We run our same $\lambda_R$ measuring apparatus on the unbinned spaxels as we do on the Voronoi-binned data. Since the unbinned data can get very noisy in the outer parts, we restrict attention in this test to \lamre. Taking only galaxies that are binned (with at least five or more spaxels per bin at $R_e$), we again find that the majority of objects have $\delta \lamre = 0$, and again we find a bias towards larger \lamre\ in the binned versus unbinned data. Again, the bias is quite small, with only 11\% of galaxies having $\lvert \delta \lamre \rvert > 0.01$ and only 5\% having $\lvert \delta \lamre \rvert > 0.02$.

\subsection{Spatial Resolution}

\begin{figure*}
\hskip 15mm
\vbox{ 
\includegraphics[width=0.4\textwidth]{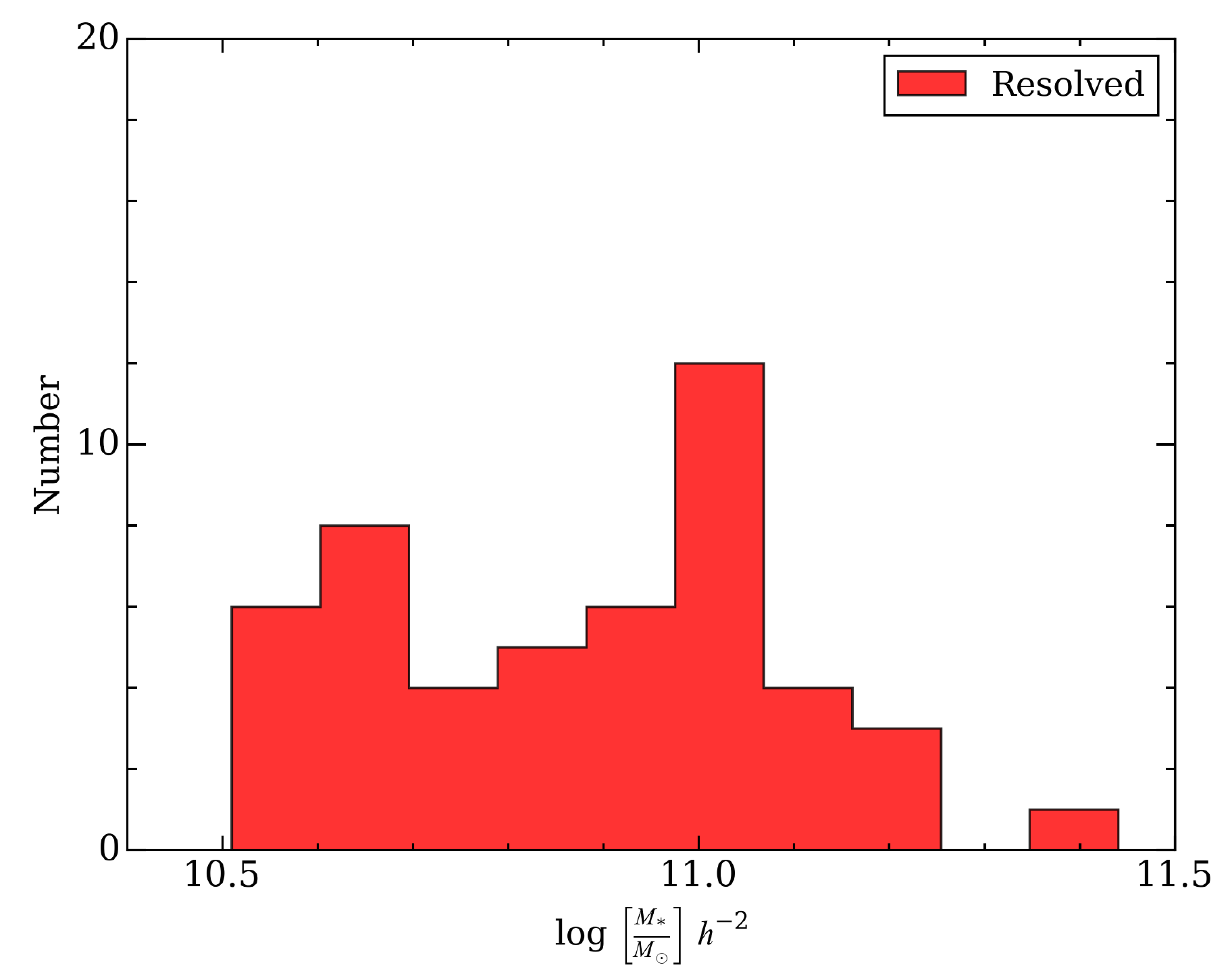}
\includegraphics[width=0.4\textwidth]{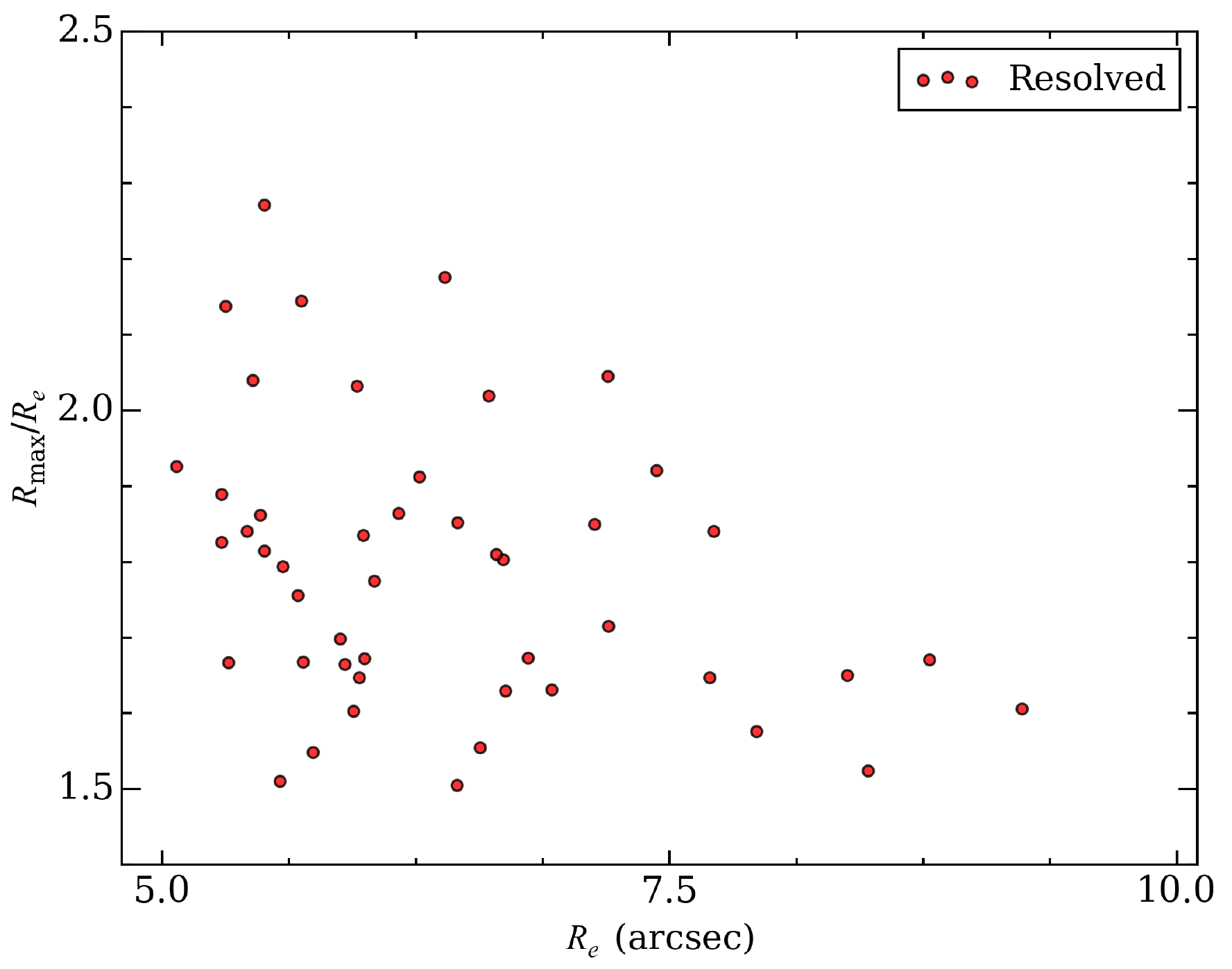}
}
\vskip -0mm
\figcaption[]{
{\it Left}: Distribution of stellar masses for our well-resolved sample that we use to model the impact of seeing. This sample includes galaxies over our full mass range.
{\it Right}: Distribution of angular size and angular extent (in $R_e$ units). The second benefit of using the MaNGA galaxies themselves to model the impact of seeing is that the coverage of these cubes extends well beyond the effective radius.
\label{fig:resolvedprop}}
\end{figure*}

Because of the design of the MaNGA survey, most galaxies are moderately resolved, with typical angular radii (in our early-type, central galaxy sample) of 5\arcsec. We cannot afford to exclude galaxies that are only marginally resolved at $R_e$ (Figure 2). In this Appendix, we explore creative ways to use the available information to derive robust $\lambda_R$ measurements that can be compared across the entire sample.

\begin{figure*}
\hskip 15mm
\vbox{ 
\includegraphics[width=0.8\textwidth]{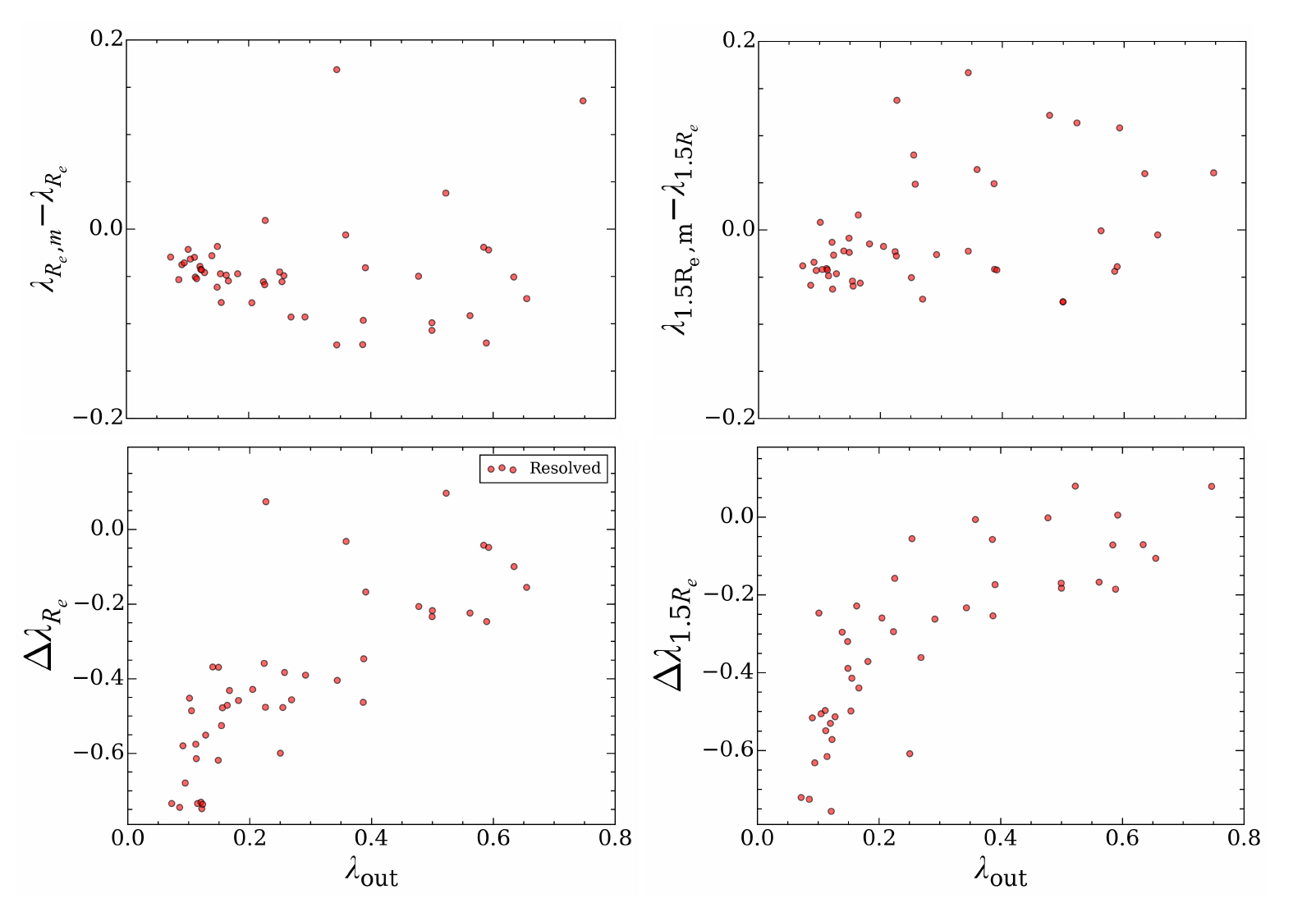}
}
\vskip -0mm
\figcaption[]{
Demonstrating how lower spatial resolution biases $\lambda_R$ using the 50 well-resolved galaxies.  In all cases, the subscript $m$ indicates the smoothed model.
{\it Left}: Top: Absolute difference ($\lambda_{R_e, m}-\lambda_{R_e}$), where $\lambda_{R_e, m}$ is recovered from the smoothed cube. Bottom: fractional difference $\Delta \lambda = (\lambda_{R_e, m}-\lambda_{R_e}$)/$\lambda_{R_e}$.
{\it Right}: Same as {\it left}, but comparing measurements at $1.5 R_e$. For $\lambda_R > 0.2$, we recover the true $\lambda_R$ value on average at $1.5 R_e$.
\label{fig:resolvedinout}}
\end{figure*}

To understand how the resolution impacts our measurements, we define a subset of 50 galaxies that have sizes $R_e > 5.5$\arcsec\ (more than four beams across $R_e$, for a median seeing FWHM of 2.5\arcsec) and radial coverage to at least $1.5 R_e$. While some previous work \citep[e.g.,][]{vandesandeetal2017} has performed simulations starting with the ATLAS$^{\rm 3D}$ sample, which has very high sampling and spatial resolution, working directly with the most spatially extended MaNGA cubes provides two key advantages. Firstly, because of the different sized MaNGA bundles, we are able to find galaxies that are both well-resolved and have coverage beyond $1.5 R_e$. Very few ATLAS$^{\rm 3D}$ cubes extend to such radii, but this is the regime where the MaNGA data are best.  Secondly, because of the small volume of ATLAS$^{\rm 3D}$, the sample is heavily weighted towards low-mass galaxies. Our well-resolved sample spans the same range in stellar mass as our parent sample (Figure \ref{fig:resolvedprop}).

Taking this sample, we degrade the spatial resolution such that the final FWHM$=R_e/3.6$ (which is typical for the sample as a whole). In degrading the data, we assume a Gaussian PSF and that the line-of-sight velocity distribution (LOSVD) for each fiber is a Gaussian, simply adopting the measured velocity and velocity dispersion as measured by pPXF for each fiber. A new LOSVD is constructed at each position by weighting each spectrum with the new PSF and combining the weighted spectra, while the new velocity and velocity dispersion are derived from a fit to the new LOSVD. We then recompute $\lambda$(R), and compare the values of $\lambda_{\rm R_e}$ between the input and smoothed cube.

In general, $\lambda_{\rm R_e}$ is sensitive to the resolution, and the situation is most severe for systems with \lamre$<0.2$. For galaxies with input \lamre\ between $0-0.2$, \lamre\ values from the simulations are typically lower by $45\%$ compared to the true values (assuming the typical MaNGA galaxy ratio of seeing to $R_e$ of $\sim 2.5$). Galaxies with \lamre$>0.2$ are impacted at the $15\%$ level on average (Figure \ref{fig:resolvedinout}, left). If we measure $\lambda$ at $1.5 R_e$ instead, the bias decreases. Galaxies with intrinsic \lamre$<0.2$ decline by only $35\%$, while galaxies with higher \lamre\ biased low by $10\%$ (with an absolute offset of $-0.02$ in $\lambda_R$). Using $\lambda$ at $1.5R_e$, however, excludes 5\% of the galaxies whose rotation curves do not quite reach this radius. Thus, we introduce a final measurement, the mean $\lambda$ as measured in the outermost 10\% of the $\lambda_R$ curve, \lamout. The quantity \lamout\ matches $\lambda$($1.5 R_e$) with no bias, and a scatter of $20\%$. To maximize the sample size, we compare objects using \lamout, but we repeat all of our analysis for $\lambda$ at $1.5 R_e$ and none of our conclusions change. 

\begin{figure*}
\hskip 15mm
\vbox{ 
\includegraphics[width=0.85\textwidth]{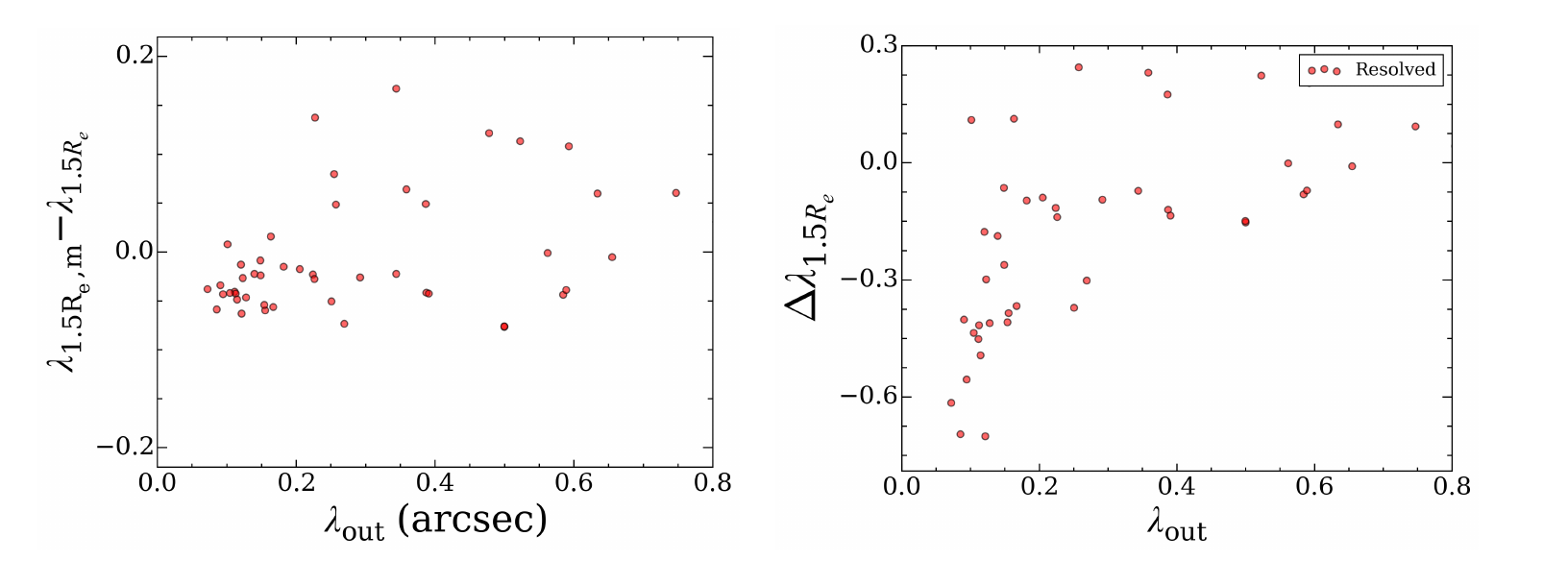}
}
\vskip -0mm
\figcaption[]{
We demonstrate how removing the central beam shifts the measured $\lambda_R$ values upward in the well-resolved galaxies after they have been smoothed, with $\Delta \lambda = (\lambda_{1.5 R_e, m}-\lambda_{1.5 R_e}$)/$\lambda_{1.5 R_e}$. The entire distribution is now consistent with no offset for \lamout$>0.2$. The large amount of scatter is partially explained by the relatively limited radial coverage relative to the model beam. 
\label{fig:lam15outhist}}
\end{figure*}

\begin{figure*}
\vbox{ 
\includegraphics[width=0.42\textwidth]{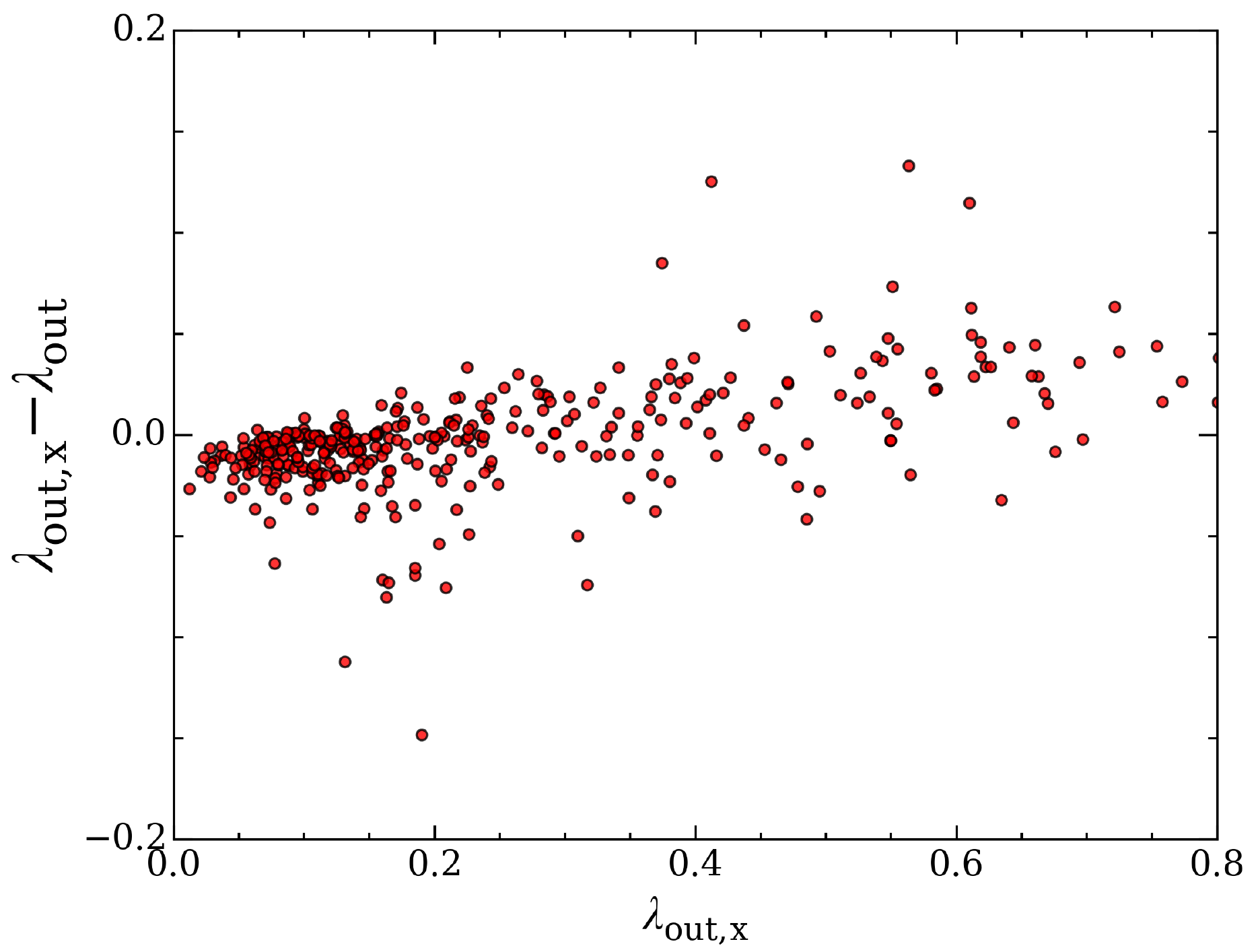}
\includegraphics[width=0.44\textwidth]{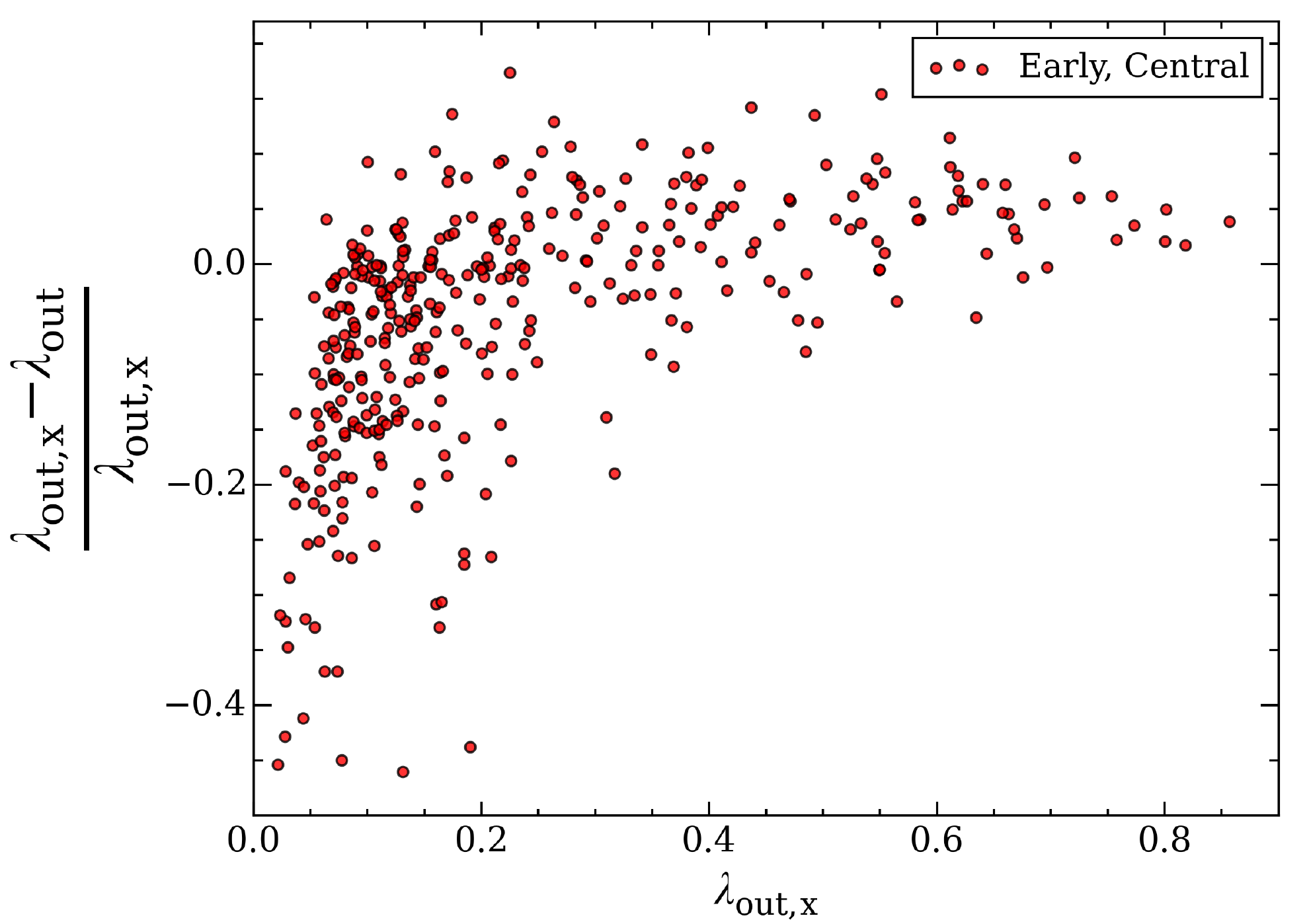}
}
\vskip -0mm
\figcaption[]{
The impact of excluding the central $2$\arcsec\ (\lamout$_\mathrm{,x}$) using the entire data set. We show the absolute difference between \lamout$_{,x}$, and \lamout\ without any exclusion (top) and the fractional difference (bottom). Note that there is a bias towards lower values for the galaxies with the lowest \lamout. Removing the central beam mitigates the 10\% bias that remains from just adopting a large-radius measurement, since the central regions have unnaturally high dispersions due to resolution effects.
\label{fig:allgalinout}}
\end{figure*}

We apply one more correction to mitigate the final 10\% bias. For obvious reasons, within the central beam ($\sim 2$ arcsec), all the rotation is measured as dispersion, naturally lowering $\lambda$ everywhere. If we excise this inner region, then both \lamout\ and $\lambda$ at $1.5R_e$ increase by $\sim 10\%$. We demonstrate this in two ways. First, we take our smoothed models, remove the central spaxels that are within the smoothed FWHM, and remeasure $\lambda$($1.5 R_e$; Fig.\ \ref{fig:lam15outhist}). We find that $\lambda$($1.5 R_e$) increases on average, such that $\Delta \lambda = (\lambda_{1.5 R_e, m}-\lambda_{1.5 R_e}$)/$\lambda_{1.5 R_e} = 0.04$. There is quite a bit of scatter introduced in these models because of their relatively limited radial coverage. As a second test, we go back to the full sample and experiment with removing the central $2$~\arcsec, in Figure \ref{fig:allgalinout}. Interestingly, the slowest rotators have their \lamout\ values drop slightly when the center is excluded (by $\sim 0.01-0.02$, which is in the noise of our \lamout\ measurements) but at \lamout$>0.2$, \lamout\ systematically increases by $\sim 10\%$. Thus we are able to fully recover the input $\lambda$($1.5R_e$) values in our simulations, on average. This corrected \lamout\ is the value utilized here.

\subsection{Impact of resolution on slow-rotator fraction}

We can use these simulations to calculate an error bar on the slow-rotator fraction, by asking how many galaxies might change designation based on residual bias in our \lamout\ values. For \lamout$>0.2$, there is effectively no bias. At lower \lamout$<0.2$, we tend to systematically underestimate \lamout\ by 30\% (Figure \ref{fig:allgalinout}). Here we estimate the resulting uncertainty in slow-rotator fraction that ensues from this uncertainty, using two methods. 

First, we take all the objects in the sample with a measured \lamout$<0.14$ (corresponding to an intrinsic \lamout$<0.2$). If we assume that these all have an intrinsic \lamout\ that is 30\% higher, we can ask what fraction of these would be considered fast rotators if we did not have a biased \lamout\ measurement for them. Of the 72 sources with \lamout$<0.14$, 16 are mis-identified as slow rotators, constituting a 10\% error in the slow-rotator fraction. The mass distribution of the low \lamout\ sources is comparable to that of the full sample, with a median $\langle$log $M_* \rangle = 10.9$~\msun. Thus, we feel confident that the residual errors incurred from our lower-than-optimal spatial resolution do not impact the main conclusions of this paper.

\begin{figure*}
\hskip 20mm
\vbox{ 
\includegraphics[width=0.7\textwidth]{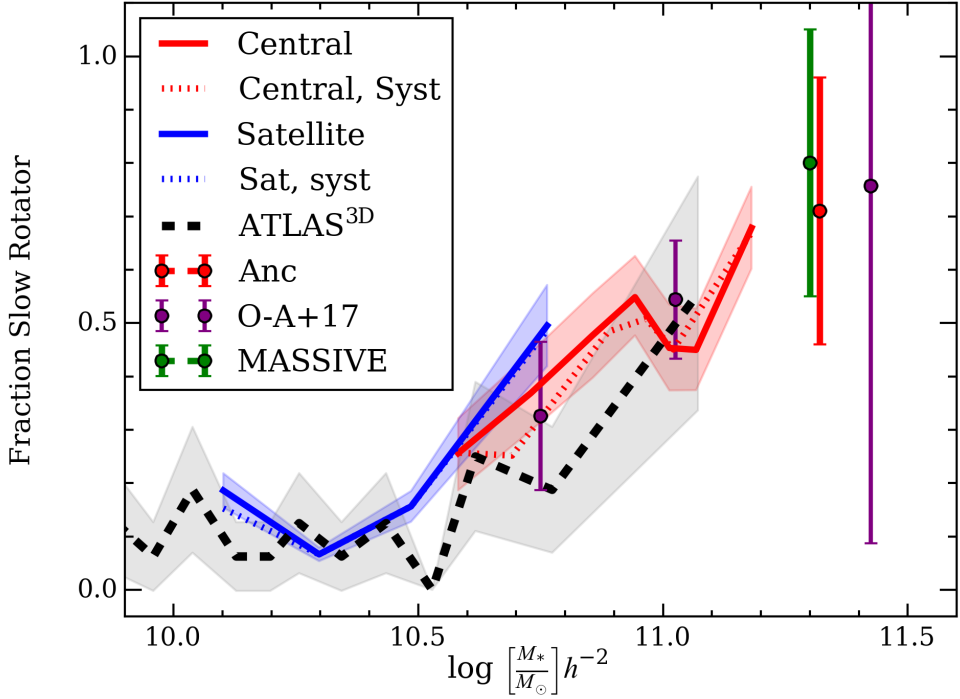}
}
\vskip -0mm
\figcaption[]{
A reproduction of Figure \ref{fig:srfrac_mstar}, including Monte Carlo estimates of the uncertainty calculated as the scatter between \lamout\ with and without the central region excluded. These Monte Carlo calculations are shown in dotted. There is not a significant difference between the measured values and these perturbed values. The spread in slow rotator fraction from the Monte Carlo runs is in all cases considerably smaller than the Poisson errors we plot, and so is not shown.
\label{fig:srfracMC}}
\end{figure*}

Second, we do an actual Monte Carlo experiment. In three bins of \lamout\ ($<0.1$,$0.1-0.2$,$>0.2$), we measure the mean offset and scatter from Figure \ref{fig:allgalinout}. In each Monte Carlo run, we perturb $\lambda/\sqrt{\epsilon}$ based on this mean offset and scatter. Based on these perturbed values, we calculate a new slow rotator fraction. We then calculate the mean and scatter in the slow rotator fraction from these runs at each mass. In Figure \ref{fig:srfracMC}, we show the perturbed values in the dotted lines; there is little difference between the dotted lines and our calculated values in solid. 

\newpage

\end{document}